\documentclass[10pt,twocolumn]{IEEEtran}
\usepackage{mathrsfs}
\usepackage{amssymb}
\usepackage{amsmath}
\usepackage{bm}
\usepackage{epsfig}
\usepackage{color}
\usepackage{csquotes}
\usepackage[font=small]{caption}
\usepackage{subcaption}
\usepackage{gensymb}
\usepackage{float}
\usepackage{bbm}
\usepackage{cite}
\usepackage{xcolor}

\newcommand\shalf{\scriptscriptstyle{1/2}}

\newtheorem{theorem}{Theorem}

\newtheorem{lemma}{Lemma}
\newtheorem{remark}{Remark}
\newtheorem{proposition}{Proposition}

\newtheorem{assumption}{Assumption}

\DeclareMathOperator*{\argmax}{arg\,max}
\usepackage{accents}
\newcommand{\ubar}[1]{\underaccent{\bar}{#1}}

\usepackage[labelformat=simple]{subcaption}

\usepackage[ruled,vlined]{algorithm2e}
\usepackage[normalem]{ulem}
\SetArgSty{textnormal}
\SetKwInput{KwResult}{Initialization}

\newcommand{\overbar}[1]{\mkern 1.5mu\overline{\mkern-1.5mu#1\mkern-1.5mu}\mkern 1.5mu}

\usepackage{cases}
\begin{document}

\title{Learning to Seek: Multi-Agent Online Source Seeking Against Non-Stochastic Disturbances}
\author{ Bin~Du,
Kun~Qian,
Christian~Claudel,
and~Dengfeng Sun
\thanks{Bin Du is with College of Automation Engineering, Nanjing University of Aeronautics and Astronautics, Nanjing 210016, China (iniesdu@nuaa.edu.cn)}
\thanks{Kun Qian and Christian Claudel are with Department of Civil, Architectural, and Environmental Engineering, the University of Texas at Austin, Austin, TX 78712, USA (\{kunqian, christian.claudel\}@utexas.edu)}%
\thanks{Dengfeng Sun is with School of Aeronautics and Astronautics, Purdue University, West Lafayette, IN 47906, USA (dsun@purdue.edu)}%
}
\maketitle

\begin{abstract}
This paper proposes to leverage the emerging~learning techniques and devise a multi-agent online source {seeking} algorithm under unknown environment. Of particular significance in our problem setups are: i) the underlying environment is not only unknown, but dynamically changing and also perturbed by two types of non-stochastic disturbances; and ii) a group of agents is deployed and expected to cooperatively seek as many sources as possible. Correspondingly, a new technique of discounted Kalman filter is developed to tackle with the non-stochastic disturbances, and a notion of confidence bound in polytope nature is utilized~to aid the computation-efficient cooperation among~multiple agents. With standard assumptions on the unknown environment as well as the disturbances, our algorithm is shown to achieve sub-linear regrets under the two~types of non-stochastic disturbances; both results are comparable to the state-of-the-art. Numerical examples on a real-world pollution monitoring application are provided to demonstrate the effectiveness of our algorithm.
\end{abstract}

\section{Introduction}\label{sec:Intro}

The problem of online source seeking, in which one or multiple agents are deployed to adaptively localize the underlying sources under a possibly unknown and disturbed environment, has gained considerably increasing attention recently among researchers in both control and robotics communities~\cite{poveda2021robust,angelico2021source,li2021source,liu2020semi}. Two challenges are of particular significance to solve such a source seeking problem: i) how to obtain a reliable perception or estimation via observations on the unknown environment; and ii) how to integrate the environment estimation with task planning for the agent(s) to seek sources in an online manner. 

In order to tackle with the above two challenges, a variety~of methodologies have been investigated in the literature, among which, the mainstream approaches are typically based on the estimation of environment gradients\mbox{\cite{ramirez2018stochastic,azuma2012stochastic,habibi2016gradient}}. Considering that the sources~are often associated with the maximum/minimum values of a function which is utilized to characterize the~state of environment, thus the gradient based approaches naturally steer the agents to search along with the direction of estimated gradients toward the locations whose gradients are close to zero. An appealing feature of this method is often attributed to the fact that only local measurements are collected during the searching process without the knowledge of agents’ global positions. However, a critical disadvantage is that the agents are easily trapped into the local extremum when the environment can not be modeled as an ideal convex/concave function. 

To further address~the above issue, recent methods,~building on certain learning techniques~\cite{rolf2020successive,du2021multi}, interplays processes of learning of an unknown environment and source seeking based on the learned environmental information. Particularly, a novel algorithm termed as~\texttt{AdaSearch} is proposed in~\cite{rolf2020successive}, which leverages the notions of upper and lower confidence bounds to guide the agent's adaptive searching for the static sources. Our previous work~\cite{du2021multi} considers a more sophisticated searching scenario, in which i) the unknown environment follows~certain linear dynamics, and thus the underlying sources are moving around; and ii) multiple agents are deployed simultaneously with the aim to cooperatively locate as many moving sources as possible. Indeed, one of the significant challenges in such a multi-agent source seeking setup is the combinatorial growth of the searching space as the increase of~the number of agents. To deal with this challenge, we~developed a novel notion of confidence bound, termed as D-UCB, which appropriately constructs a polytope confidence set and~helps decompose the searching space for each agent. As a consequence, a linear complexity is achieved for our algorithm with respect to the number of agents, which enables~the computation-efficient cooperation among the multiple agents.

Despite the remarkable feature of our D-UCB algorithm in reducing the computational complexity, one critical drawback is its dependence on the precise knowledge of the environment dynamics. {Nevertheless, considering that uncertainties and/or disturbances are almost ubiquitous in practice, the knowledge of an exact model on the environment is barely available~when considering real-world applications. To take into account the disturbances in system dynamics, a set of classical \mbox{approaches}, such as linear quadratic regulator, incorporates~the stochastic processing noise which is usually assumed to be independent and identically (Gaussian) distributed and in most cases with zero-mean. Recently, with the great advancement~of learning theory applied into control problems, relevant works started to turn to a new paradigm where the stochastic~disturbances are replaced by~\mbox{non-stochastic} ones. It is well recognized~that, in most problems, the non-stochastic setup is more challenging than the stochastic one, as the standard statistical properties of the disturbances are no longer available. In addition, it~is also more general, on the~other hand, since the non-stochastic disturbances can not only characterize the modeling deviation of the environment, but also be interpreted as the one which is arbitrarily injected by an underlying adversary.} As such, we consider~in~this~paper the multi-agent online source seeking problem with the non-stochastic setup where the environment is disturbed by two types of non-stochastic disturbances. Our objective is to enhance the D-UCB algorithm with the~capability  of dealing with the non-stochastic disturbances while still enjoying the low computational complexity with a guaranteed source seeking performance.

\subsection{Related Works}

As mentioned earlier, the predominate approaches to solve the source seeking problem, including the well-known technique of \textit{extremum seeking control}~{\cite{feiling2018gradient,dougherty2016extremum}}, often build on the environment gradient estimation. These approaches can be indeed viewed as variants of  the first-order optimization~algorithm, which drive the agent to search for the local extremum values. {In particular, by modeling the unknown environment as a time-invariant and concave real-valued function, the authors in~\cite{li2014cooperative} designed the distributed source seeking~control~law~for a group of cooperative agents. Besides, the diffusion process is further considered in~\cite{fabbiano2014source} for investigating the scenario~of dynamical environments. The source seeking problems are~also studied in~\cite{brinon2015distributed,fabbiano2016distributed} by forcing the multiple agents to~\mbox{follow} a specific circular formation. In addition,~the~stochastic~\mbox{gradient} based methods are proposed in~\cite{atanasov2012stochastic,atanasov2015distributed} when considering that the gradient estimation is subject to environment and/or measurement noises.} We should note that, also inherited from the first-order optimization algorithm, the above gradient based methods are very likely to be stuck at local extremum~points when the considered environment is non-convex/non-concave. Furthermore, the gradient estimation is also sensitive~to the measurement/environment noise, and thus additional statistical properties on the noise, such as known distribution with zero-mean, need to be imposed as assumptions in the problem setup.

Whilst it is unknown how to deal with noises without~statistical properties in the context of source seeking, non-stochastic disturbance has been considered increasingly broadly in control communities. Within the classical robust control framework, the non-stochastic disturbance is often~treated by considering the worst-scenario performance, see e.g.,~\cite{zhou1998essentials}. However, more recent works related~to~the learning based control mainly concern about the development of adaptive approaches which aim at controlling typically a linear system with adversarial disturbances while optimizing certain objective function with respect to the system states and control inputs\mbox{\cite{agarwal2019online,foster2020logarithmic,simchowitz2020improper,pmlr-v117-hazan20a,simchowitz2020making}}.~To measure the performance of adaptive controllers, the~notion of regret is adopted; that is,
to measure the discrepancy between the gain/cost of the devised controller and that of the best
one in hindsight. In particular, the authors in~\cite{agarwal2019online} devise the first $\mathcal{O}(\sqrt{K})$-regret algorithm by assuming a convex cost function and known system dynamics. Afterwards, such a regret bound is enhanced to be logarithmic in~\cite{foster2020logarithmic,simchowitz2020improper} within the same problem setup. To further relax the requirement of the known dynamics, the authors in~\cite{pmlr-v117-hazan20a} develop the algorithm which attains $\mathcal{O}(K^{2/3})$-regret, and such a bound is also improved to $\mathcal{O}(\sqrt{K})$ later in~\cite{simchowitz2020improper,simchowitz2020making}.~Though the above works have investigated quite thoroughly the non-stochastic setting in the context of learning based control, we remark that our paper considers a different problem where some standard conditions in control, such as~controllability and observability, can be no longer simply assumed. In fact, our problem is more~related~to a sequential decision process; that is, the agents make their source seeking decisions in sequence while interplaying with perception of the unknown environment.

This sequential feature also makes our setting closely related to the well-known problem of multi-armed bandits. Therefore, another rich line of relevant works is on the series of bandit algorithms. More specifically, involved with the \mbox{non-stochastic} disturbances, linear bandits are investigated within two~\mbox{settings} of non-stationary environment and adversarial corruptions,~respectively. While the former one interprets the non-stochastic disturbance as a variation of the environment, the latter one~is corresponding to corruptions injected by potential adversaries. Both cases are well studied in literature with the development of algorithms guaranteeing sub-linear regrets. To deal with the environmental non-stationarity, the WindowUCB~algorithm is 
first proposed in~\cite{cheung2019learning} along with the technique of sliding-window least squares. It is shown that the algorithm achieves the regret of $\widetilde{\mathcal{O}}(K^{2/3}B_K^{1/3})$ where $B_K$ is a measure to the~level of non-stationarity. The same regret is proved for the weighted linear bandit algorithm proposed in~\cite{russac2019weighted}, which leverages on the weighted least-square estimator. Further, a simple restart strategy is developed in~\cite{zhao20a}, obtaining the same regret. It is indeed proved that the $\widetilde{\mathcal{O}}(K^{2/3}B_K^{1/3})$-regret~is~the optimal one that can be achieved in the setting of non-stationary bandits. In terms of the adversarial bandits, a robust algorithm is proposed in~\cite{ding2022robust} which guarantees the $\widetilde{\mathcal{O}}(B_K{K}^{3/4})$-regret, and thus it is sub-linear only if the level of adversarial corruptions satisfies $B_K = o(K^{1/4})$. More recently, such a regret has~been improved to $\widetilde{\mathcal{O}}(B_K+\sqrt{K})$ in~\cite{bogunovic2021stochastic,he2022nearly} which is also shown to be nearly optimal in the adversarial setting. It can be concluded from the above discussion that, once $B_K$ grows~sub-linearly, the regrets in both cases are guaranteed to be sub-linear. These are also the state-of-the-art that we are expected to achieve for our algorithm to be developed in this work.

\subsection{Statement of Contributions}
This paper proposes an online source seeking algorithmic framework using the emerging learning technique, which is capable of i) dealing with the unknown environment in the presence of non-stochastic disturbances; and ii) taking advantages of the cooperation among the multi-agent network.~In~terms of the non-stochastic disturbances, two specific types of them are considered: i) an external one which disturbs the measurable states of the environment; and ii) an internal one which is~truly evolved with the environment dynamics. To deal with them, an unified technique of discounted Kalman filtering is proposed to estimate the unknown environment~states while mitigating the disturbances. Meanwhile, to build the cooperation among multiple agents and avoid the combinatorial complexity, we leverage the polytope confidence set, and as a result, the proposed algorithm is exceptionally computation-efficient in the multi-agent setting. It is shown by the regret analysis that our algorithm attains sub-linear regrets against both types of non-stochastic disturbances. The obtained two regrets are both comparable to the state-of-the-art in the studies of non-stationary and adversarial bandit algorithms. At last, all theoretical findings are validated by simulation examples on a real-world pollution monitoring application.

\section{Problem Statement}


\subsection{Unknown Environment with Non-Stochastic Disturbances}
Consider an obstacle-free environment which is assumed to be bounded and discretized  by a finite set of points $\mathcal{S}$ where each $\mathbf{s} \in \mathcal{S}$ represents the corresponding position. Suppose~that the unknown state of the environment at each discrete time $k$ is described by a real-valued function $\phi_k(\cdot): \mathcal{S} \to \mathbb{R}_+$ which maps the positional information $\mathbf{s}$ to a positive quantity $\phi_k(\mathbf{s})$ indicating the environmental value of interest. Let us denote~$N$ the total number of all points, i.e., $N = |\mathcal{S}|$, and for simplicity, denote $\bm{\phi}_k \in \mathbb{R}_+^N$ the vector which stacks all individual $\phi_k(\mathbf{s})$. Further, to characterize dynamics~of~the changing environment, we consider that the evolution of state $\bm{\phi}_k$ is basically governed by the following nominal linear time-varying (LTV) model
\begin{align}\label{linearModel}
  \bm\phi_{k+1} = A_{k+1} \bm\phi_{k},
\end{align} 
where the state transition matrix $A_k \in \mathbb{R}^{N \times N}$ is assumed to be known a prior. In order for the considered source seeking problem to be well-defined, we need the state $\bm{\phi}_k$ to be neither explosive nor vanishing to zero, which can be ensured by the following assumption.
\begin{assumption}\label{assump:dynamicsBound}
  For the LTV dynamics~\eqref{linearModel}, there exists a pair of uniform lower and upper bounds $0 < \ubar{\alpha}\le \bar{\alpha} < \infty $ such that, for $\forall k \ge t > 0$,
  \begin{align}
    \ubar{\alpha} \cdot \mathbf{I}_N \preceq A[k:t]^\top A[k:t] \preceq \bar{\alpha} \cdot \mathbf{I}_N,
  \end{align}
  where $\mathbf{I}_N$ represents the $N \times N$ identity matrix and the state propagation matrix\footnote{By convention, we let $A[k:t] = \mathbf{I}_N$ when $k < t$.} is defined as $ A[k:t]:= A_k A_{k-1} \cdots A_t $.
\end{assumption}

We should note that the above Assumption~\ref{assump:dynamicsBound} not only helps confine the behavior of the environment states, but also implies the invertibility of the state transition matrices $A_k$'s which~aids the subsequent regret analysis of our algorithm. In fact, such an assumption is not unusual in the study of system control and estimation problems; see e.g., \cite{li2019boundedness,battistelli2014kullback,battistelli2014consensus,cattivelli2010diffusion}.

Now, in order to further impose the underlying disturbances into the environment model, let us consider the following two types of non-stochastic ones on top of the nominal dynamics:
\begin{subnumcases}
	{\label{disturb}\hspace{-50pt}}
		\texttt{Type I}: &\quad $\widetilde{\bm\phi}_{k+1} = A_{k+1} {\bm\phi}_{k} + \bm{\delta}_k$.\label{typeI_disturb}\\
		\texttt{Type II}:&\quad $\widetilde{\bm\phi}_{k+1} = A_{k+1} \widetilde{\bm\phi}_{k} + \bm{\delta}_k$,\label{typeII_disturb}
\end{subnumcases}
Note that in both types, $\widetilde{\bm\phi}_k\in \mathbb{R}_+^N$ denotes the~disturbed~state. However, while the first type of disturbance can be interpreted as an \textit{external} one since $\bm{\phi}_k$ in~\eqref{typeI_disturb} is still evolved according to the nominal dynamics~\eqref{linearModel} and the disturbance $\bm{\delta}_k$ only affects the state $\widetilde{\bm\phi}_{k+1}$ in one step, the second type can be viewed as an \textit{internal} one since the disturbance $\bm{\delta}_k$ is intrinsically~imposed into the dynamics and accumulated during the evolution of~$\widetilde{\bm\phi}_k$. In fact, we shall remark that the two types of disturbances both find a wide range of real-world applications.
For instance, in the scenario of pollution monitoring which is investigated in our simulations, the external disturbance could correspond to certain unrelated emitters which will not change the locations of sources of interest but interfere the perceptible environment states, the internal one might result from some environmental conditions, such as wind, which will truly affect the diffusion of pollutants and thus change their positions.
It is also enlightened by the provided example that the localization of sources should be considered differently for the above two cases. More details will be found in Section~\ref{subsec:SourceSeeking}. In addition, we note that the internal disturbance can be also used to capture to some extent the unmodeled dynamics of the unknown environment. However, no matter which type of disturbances is involved in the process, only the disturbed state $\widetilde{\bm\phi}_k$ is measurable for the agents which are employed to operate in the environment later.

As we have remarked earlier, the {disturbances} of both types are supposed to be non-stochastic, i.e., no statistical property in any form is assumed regarding $\bm{\delta}_k$. Instead, to characterize the effect of both disturbances in long term, we consider to impose the following assumption.
\begin{assumption}\label{assump:disturbBound}
 There exists a positive sequence $\{B_K\}_{K \in \mathbb{N}_+}$ such that, for $\forall K \ge 0$,
  \begin{align}
    \sum_{k=0}^K \| \bm{\delta}_k\| \le B_K.
  \end{align}
\end{assumption}

\begin{remark}
	The sequence $\{B_K\}_{K \in \mathbb{N}_+}$ in Assumption~\ref{assump:disturbBound} is not necessarily required to be bounded by some constant~in~our work. In fact, we consider the problem under the condition~that $B_K$ increases at a sub-linear rate and aim to provide a~performance guarantee for our algorithm on the dependence of~$B_K$. It is often implied by the sub-linear increasing $B_K$ that~either the total number of occurrence of the disturbance $\bm\delta_k$ increases sub-linearly or the effect of disturbance $\|\bm\delta_k\|$ vanishes to zero over the  time-steps $k$. While the former is often referred to as the abrupt-changing disturbance, the latter is regarded as the slowly-varying one. In addition, in the context of learning theory in adversarial/non-stationary settings, such a sequence
	$\{B_K\}_{K \in \mathbb{N}_+}$ is also viewed as an attack budget of an adversary; see e.g.,~\cite{yang2020adversarial,bogunovic2021stochastic}. 
\end{remark}

\subsection{Multi-Agent Source Seeking}\label{subsec:SourceSeeking}

With the aim to locate the potential sources which usually correspond to the extreme values in the unknown environment state, we deploy a network of $I$ agents and expect each of them $i \in \mathcal{I}:=\{1,2,\cdots,I\}$ to seek its best positions~$\mathbf{p}_k^\star[i] \in \mathcal{S}$ at each time $k$ by solving the following~maximization problem,
\begin{equation}\label{basicDCM}
\begin{aligned}
\mathop {\text{maximize} }\limits_{\mathbf{p}[i] \in \mathcal{S},\, i \in \mathcal{I}}  \;\; &F_k(\mathbf{p}[1],\mathbf{p}[2],\cdots, \mathbf{p}[I]) = \hspace{-5pt} \sum_{\mathbf{s} \in \cup_{i=1}^I\mathbf{p}[i]} \phi_k(\mathbf{s}).
\end{aligned}
\end{equation}
Notice that the summation involved in the objective function $F_k(\cdot) : \mathcal{S}^I \to \mathbb{R}_+$ takes into account the \textit{union} of positions $\mathbf{p}[i]$'s, therefore all agents will naturally  tend to locate as many \textit{distinct} positions as possible for the purpose of maximizing $F_k(\cdot)$. In addition, it is now clear to see the reason why Assumption~\ref{assump:dynamicsBound} would be needed, i.e., the maximization in~\eqref{basicDCM} is  otherwise not well-defined if the environment state $\bm{\phi}_k$ explodes or vanishes to zero. Further, we should also note that an inherent difference will take place in the counted state $\bm{\phi}_k$ involved in the objective function when considering the disturbances of the~two types. More precisely, for the first type of disturbance, i.e., the external one, the positions of sources should be indeed reflected by the undisturbed $\bm\phi_k$, though only the information of disturbed $\widetilde{\bm\phi}_k$ is measurable for agents. On the contrary, for the second type, i.e., the internal disturbance, the disturbed $\widetilde{\bm\phi}_k$ should be taken into account in~\eqref{basicDCM}, since $\bm\delta_k$ is evolved in the environment dynamics and changes the positions of sources. On this account, we emphasize that while the maximization problem~\eqref{basicDCM} is precisely the one that the agents would like to solve when considering the external disturbance, yet for the internal one, the objective function should be amended as
\begin{align}\label{objFunc_II}
	\widetilde{F}_k(\mathbf{p}[1],\mathbf{p}[2],\cdots, \mathbf{p}[I]) = \hspace{-5pt} \sum_{\mathbf{s} \in \cup_{i=1}^I\mathbf{p}[i]}\widetilde{\phi}_k(\mathbf{s}).
\end{align}

 With the above difference presented in the objective functions, the main challenges in solving the maximization problems are also distinguishable in principle. Whilst the former is to extract the true information hidden in $\bm\phi_k$ in the case that only $\widetilde{\bm\phi}_k$ is accessible, the latter is to identify and compensate the unmodeled disturbance $\bm\delta_k$. Despite this difference, we develop in this paper an unified algorithmic framework for both cases, enabling the agents to track the dynamical sources in an online manner. We remark that this is also one of the main contributions of our work.

Another common technical issue, regardless of types of the disturbances involved, is to deal with the estimation of the environment. Therefore, we leverage on the following linear stochastic measurement model,
\begin{align}\label{dynMeasurement}
  \mathbf{z}^i_k = H^i\big(\mathbf{p}_k[i]\big) \widetilde{\bm{\phi}}_k + \mathbf{n}_k^i,
\end{align}
where $\mathbf{z}_k^i\in \mathbb{R}^{m}$ is the $i$-th agent's obtained measurement at the time-step $k$; $H^i\big(\mathbf{p}_k[i]\big) \in\mathbb{R}^{m\times N}$ denotes the measurement matrix depending on the agent's position $\mathbf{p}_k[i]$; and $\mathbf{n}_k^i \in \mathbb{R}^{m}$ is the measurement noise which is assumed to be independent and identically distributed (\textit{i.i.d.}) Gaussian with zero mean and variance  $V^i = v^i\cdot \mathbf{I}_m$. We shall note that the measurement matrix $H^i\big(\mathbf{p}_k[i]\big)$ is not specified in~\eqref{dynMeasurement}. In fact, it can be~defined by various means based on the agent's position. Nevertheless, we assume that each $H^i\big(\mathbf{p}_k[i]\big)$ has the following basic form,
\begin{align}\label{MeasureMatrix}
  H^i\big(\mathbf{p}_k[i]\big) = \big[\mathbf{e}_l\big]^\top_{l \in \mathcal{C}_k^i},
\end{align}
where $\mathbf{e}_l$ denotes the unit vector, i.e., the $l$-th column of the identity matrix, and $\mathcal{C}_k^i$ is the set of positions which are covered by the agent's sensing area at the time-step $k$. It is natural to assume that the position where the agent currently locates falls into its sensing area, i.e., $\mathbf{p}_k[i] \in \mathcal{C}_k^i$.

\section{Development of the Online ALgorithm}\label{sec:onlineAlgo}

In this section, we develop our online source seeking algorithm which relies on two central ingredients: 1) a \textit{discounted Kalman filter}, which is capable of providing an estimation on the unknown environment while dealing with the two types of non-stochastic disturbances in an unified framework; and 2) a \textit{D-UCB} approach, which helps determine the agents' seeking positions sequentially in an computation-efficient manner.
 
\subsection{Estimation of the Environment States with Disturbances}
According to the measurement model~\eqref{dynMeasurement} introduced in the previous section, let us first express it as a compact form which counts for all agents within the network. For this purpose, we stack all the measurements $\mathbf{z}_k^i$'s and also the noise $\mathbf{n}_k^i$'s as the concatenated vectors $\mathbf{z}_k \in \mathbb{R}^M$ and $\mathbf{n}_k \in \mathbb{R}^M$ with $M=mI$, e.g., $\mathbf{z}_k := [(\mathbf{z}^1_k)^\top, (\mathbf{z}^2_k)^\top, \cdots, (\mathbf{z}^I_k)^\top]^\top \in \mathbb{R}^M$. Likewise, we define the concatenated measurement matrix $H_k \in \mathbb{R}^{M \times N}$ by stacking all local $H^i(\mathbf{p}_k[i])$'s. Consequently, the measurement model of the compact form can be written as
\begin{align}\label{compactMeasurement}
  \mathbf{z}_k = H_k\widetilde{\bm{\phi}}_k + \mathbf{n}_k.
\end{align}
Note that, in the notation $H_k$, we have absorbed for simplicity the dependency on agents' positions $\mathbf{p}_k[i]$'s into the index~$k$. In addition, by our assumption on the measurement noise, one can have that the concatenated noise $\mathbf{n}_k$ is also \textit{i.i.d.} Gaussian with zero-mean and variance being
\begin{align}\label{covMat}
  V := \text{Diag}\{V^1, V^2, \cdots,V^I\} \in \mathbb{R}^{M \times M}.
\end{align}

Equipped with the agents' measurement model in its compact form~\eqref{compactMeasurement}, we are now ready to present the technique of discounted Kalman filtering. Similar to the standard Kalman filter, we also use mean $\widehat{\bm{\phi}}_k \in \mathbb{R}^N$ and covariance $\Sigma_k \in\mathbb{R}^{N\times N}$ to recursively generate estimates of the unknown environment. However, a primary difference is that two positive sequences of weights $\{\lambda_k\}_{k \in \mathbb{N}_+}$ and $\{\omega_k\}_{k \in \mathbb{N}_+}$ are imposed in the filtering process with the aim to mitigate the effect of disturbances presented in the environment. Keep this in mind, the discounted Kalman filter performs the following recursions,
\begin{subequations}\label{Kalman}
  \begin{align}
    &\Sigma_{k+\shalf} = \big(\Sigma_{k}^{-1} + \lambda_kY_k \big)^{-1},\label{KalmanSigmahalf}\\
    &\widehat{\bm{\phi}}_{k+\shalf} = \widehat{\bm{\phi}}_{k} + \lambda_k\Sigma_{k+\scriptscriptstyle{1/2}}(\mathbf{y}_k - Y_k\widehat{\bm{\phi}}_{k})\label{KalmanPhihalf},\\
    &\Sigma_{k+1} = A_{k+1}\big(\Sigma_{k+\shalf}^{-1}+(\omega_k - \omega_{k-1})\Gamma_k^{-1}\big)^{-1}A_{k+1}^\top,\label{KalmanSigma}\\
    &\widehat{\bm{\phi}}_{k+1} =\Sigma_{k+1}A_{k+1}^{-\top}\Sigma_{k+\shalf}^{-1}\widehat{\bm{\phi}}_{k+\shalf},\label{KalmanPhi}\\
    &\Gamma_{k+1} = A_{k+1} \Gamma_{k}A_{k+1}^{\top}.\label{KalmanDynamics}
  \end{align}
 \end{subequations} 
Notice that $\Sigma_{k+\shalf}\in\mathbb{R}^{N\times N}$ and $\widehat{\bm{\phi}}_{k+\shalf}\in\mathbb{R}^{N}$ here~denote the~intermediate results during the recursions; $\Gamma_{k}\in\mathbb{R}^{N\times N}$ is an auxiliary matrix initialized by $\Gamma_0 = \mathbf{I}_N$; and the variables \mbox{$\mathbf{y}_k:=H_k^\top V^{-1}\mathbf{z}_k \in \mathbb{R}^N$} and $Y_k := H_k^\top V^{-1}H_k \in\mathbb{R}^{N\times N}$, which can be readily acquired by consensus schemes, i.e.,~\cite{olfati2005consensus}, incorporate latest measurements into the update of estimates. Next, to better show how the imposed weights help deal with the non-stochastic disturbances, we present in the subsequent lemma another expression of the discounted Kalman filter~\eqref{Kalman}.

\begin{lemma}\label{lemma:estRecursion}
  Suppose that the state estimates $\widehat{\bm{\phi}}_k$ and $\Sigma_k$ are generated by \eqref{Kalman} with the initialization and $\omega_{-1} = 0$, then at each iteration $k$, it is equivalent to have
  \begin{subequations}\label{Kalman_recursion}
    \begin{align}
      \Sigma_k &= A[k:1]\Upsilon_k^{-1}A[k:1]^{\top};\label{Kalman_recursion_Sigma}\\
      \widehat{\bm{\phi}}_k &= A[k:1]\Upsilon_k^{-1}\Big(\Sigma_0^{-1} \widehat{\bm{\phi}}_0 + \sum_{t=0}^{k-1}\lambda_tA[t:1]^\top \mathbf{y}_t\Big),\label{Kalman_recursion_phi}
    \end{align}
   \end{subequations}
   where the matrix $\Upsilon_k \in \mathbb{R}^{N \times N}$ is defined as
  \begin{align}\label{upsilon}
    \Upsilon_k := \Sigma_0^{-1} + \sum_{t=0}^{k-1}\lambda_tA[t:1]^\top Y_t A[t:1] + \omega_{k-1} \cdot\mathbf{I}_N.
  \end{align}
\end{lemma}

\begin{IEEEproof}
  See Appendix I.
\end{IEEEproof}

\vspace{5pt}
\begin{remark}
  According to the form~\eqref{Kalman_recursion} of the discounted Kalman filter, it can be observed that the sequence $\{\lambda_k\}_{k \in \mathbb{N}_+}$ serves to adjust weights on the measurements obtained during the process. Considering that the cumulative quantity of the disturbance is upper bounded by the sequence $\{B_K\}_{K \in \mathbb{N}+}$; see Assumption~\ref{assump:disturbBound}, this implies that, in general, the influence of disturbances vanishes over time if $B_K$ increases sub-linearly. In this case, the significant disturbance which took place at the early stage can be expected to be gradually mitigated by performing the discounted Kalman filtering. Further, unlike the weight $\lambda_k$ which is only added to the measurements locally, another sequence of weights $\{\omega_k\}_{k \in \mathbb{N}_+}$ is applied to globally adjust the covariance $\Sigma_k$ so that it can compensate the effect of internal disturbances more directly.
\end{remark}


\subsection{Multi-Agent Online Source Seeking via D-UCB}

 Based on $\widehat{\bm{\phi}}_k$ and $\Sigma_k$, we now introduce the key notion of D-UCB $\bm{\mu}_k \in \mathbb{R}^N$, which is defined as follows,
 \begin{align}\label{UCB}
  \bm{\mu}_k: = \widehat{\bm{\phi}}_k + \beta_k(\delta) \cdot\text{diag}^{1/2}(\Sigma_k).
 \end{align}
 Note that the operator $\text{diag}^{1/2}(\cdot): \mathbb{R}^{N\times N} \to \mathbb{R}^{N}$ maps the square root of the matrix diagonal elements to a vector, and the sequence $\{\beta_k(\delta)\}_{k \in \mathbb{N}_+}$ depending on a predefined confidence level $\delta$ will be specified subsequently in the next section. With the aid of the defined D-UCB $\bm\mu_k$, one can update the agents' seeking positions in an online manner, by solving the following maximization problem:
\begin{align}\label{algo}
  \mathbf{p}_{k} \in \argmax_{\mathbf{p}[i] \in \mathcal{S},\, i \in \mathcal{I}} \sum_{\mathbf{s} \in \cup_{i=1}^I\mathbf{p}[i]} {\mu}_k(\mathbf{s}).
\end{align}
Here, $\mathbf{p}_k \in \mathcal{S}^I$ stacks the decided seeking positions $\mathbf{p}[i]_k$'s for all agents, and likewise, ${\mu}_k(\mathbf{s}) \in \mathbb{R}$ represents one component of the vector~${\bm{\mu}}_k$ which corresponds to the position $\mathbf{s} \in \mathcal{S}$. The complete multi-agent online source seeking scheme under the environment with disturbances is outlined in Algorithm~\ref{algo:UCB}.

\begin{algorithm}
 \SetAlgoLined
 \caption{Multi-agent online source seeking under environment with non-stochastic disturbances}
\vspace{5pt}
  \KwResult{\parbox{\dimexpr\textwidth+2\algomargin\relax}{Each agent $i$ initializes its estimates $\widehat{\bm\phi}^0$}\\
  \parbox{\dimexpr\textwidth-2\algomargin\relax}{and $\Sigma_0$, and computes its initial position $\mathbf{p}_0[i]$. Set the}\\
    \parbox{\dimexpr\textwidth-2\algomargin\relax}{confidence $\delta$ and generate the sequence $\{\beta_k(\delta)\}_{k \in \mathbb{N}_+}$. }\\
  } 
  \vspace{5pt}
  \While{the stopping criteria is NOT satisfied}{
  \vspace{5pt}
  Each agent $i$ \textbf{simultaneously} performs\\
  \vspace{5pt}
  {\texttt{\textbf{Step 1}}} (\textit{Measuring}): Obtain the measurement $\mathbf{z}_i^k$ based on the measurement matrix $H^i(\mathbf{p}_k[i])$;\\
  \vspace{5pt}

   {\texttt{\textbf{Step 2}}} (\textit{Discounted Kalman Filtering}): Collect information from neighbors, \hspace{-2pt}obtain the estimates $\widehat{\bm{\phi}}_{k+1}$ and $\Sigma_{k+1}$ by~\eqref{Kalman};
   \vspace{5pt}

   {\texttt{\textbf{Step 3}}} (\textit{D-UCB Computing}): Compute via~\eqref{UCB} the updated D-UCB $\bm{\mu}^{k+1}$ by $\widehat{\bm{\phi}}_{k+1}$ and $\Sigma_{k+1}$;
   \vspace{5pt}

   {\texttt{\textbf{Step 4}}} (\textit{Seeking Positions Updating}): Assign the new seeking position $\mathbf{p}_{k+1}[i]$ by solving~\eqref{algo}.
      \vspace{5pt}

  Let $k \leftarrow k+1$, and continue.
   }

   \label{algo:UCB}
\end{algorithm}

\section{Regret Analysis}

In this section, we provide theoretical performance guarantee for our algorithm by the notion of regret. More specifically, we perform the regret analysis for both cases which are subject to the two types of non-stochastic disturbances, respectively. By showing the sub-linear cumulative regrets for both cases, it is ensured that the agents are capable of tracking the dynamical sources under an unknown and disturbed environment.

\subsection{On the Disturbance of Type I}\label{subsec:regretAnalysis_I}

As we have informed in the previous discussion, for the first type of disturbance, the objective function $F_k(\cdot)$ in~\eqref{basicDCM} takes into account the undisturbed state $\bm{\phi}_k$. Therefore, we introduce the notion of regret for the first case as follows,
\begin{align}
	r_k := F_k(\mathbf{p}_k^\star) \hspace{-2pt}-\hspace{-2pt} F_k(\mathbf{p}_k),
\end{align}
where $\mathbf{p}_k^\star$ denotes the optimal solution to problem~\eqref{basicDCM} and $\mathbf{p}_k$ corresponds to the decision generated by our source seeking algorithm. Here, we aim to show that the cumulative regret, i.e., $R_K \hspace{-2pt}:= \sum_{k=0}^K \hspace{-2pt}r_k$, increases sub-linearly with respect to the number of time-steps $K$, namely the regret $r_k$ converges to zero on average. To this end, let us first show the following result which formalizes that the D-UCB $\bm\mu_k$ provides indeed a valid upper bound for the unknown state $\bm\phi_k$.

\begin{proposition}\label{prop:DUCB_I}
	Under Assumptions~\ref{assump:dynamicsBound} and~\ref{assump:disturbBound}, let $\widehat{\bm{\phi}}_k$ and $\Sigma_k$ be generated by the discounted Kalman filter~\eqref{Kalman} with $\omega_k \equiv 0$ and $\lambda_k = \min\{1, \bar{\lambda}/\|Y_k\|_{\Sigma_k}\}$. Suppose that the initialization satisfies $\ubar{\sigma}\cdot\mathbf{I}_N\preceq\Sigma_0\preceq\bar{\sigma}\cdot\mathbf{I}_N$ and likewise the noise variance has $\ubar{v}\cdot\mathbf{I}_M\preceq V\preceq\bar{v}\cdot\mathbf{I}_M$, then it holds that, 
  \begin{align}\label{IneqUCB_I}
    {\mathbb{P}( \bm{\phi}_k \;{\preceq}\; \bm\mu_k ) \ge 1-\delta}, \quad \forall k \ge0,
  \end{align}
  where ${\preceq}$ is defined in element-wise, the probability $\mathbb{P}(\cdot)$ is taken on random noises $(\mathbf{n}_1, \mathbf{n}_2, \cdots,\mathbf{n}_k)$ and the sequence $\{\beta_k(\delta)\}_{k\in\mathbb{N}_+}$ in D-UCB is chosen satisfying
  \begin{equation}\label{betaK_I}
  \begin{aligned}
    \beta_k(\delta) \ge& \sqrt{N} \cdot \Bigg(\bar{\lambda} B_k+C_1 \\
    &+C_2\sqrt{N}\cdot \sqrt{\log\Big(\frac{\bar{\sigma}/\ubar{\sigma} + \bar{\alpha}\bar{\sigma}\cdot k / \ubar{v}^2}{\delta^{2/N}}\Big)}\Bigg),
  \end{aligned}
  \end{equation}
  where $B_k$ is defined in Assumption~\ref{assump:disturbBound}, $C_1 = \|\widehat{\bm{\phi}}_0 - \bm{\phi}_0\| /\sqrt{\ubar{\sigma}}$ and $C_2 = \bar{v}^2\sqrt{\max\{2,2/\ubar{v}\}}$.
\end{proposition}

\vspace{5pt}
\begin{IEEEproof}
  See Appendix II-A.
\end{IEEEproof}
\vspace{5pt}

It can be concluded from Proposition~\ref{prop:DUCB_I} that the D-UCB $\bm\mu_k$ is guaranteed to be an upper bound for $\bm\phi_k$ with probability at least $1-\delta$. In fact, considering that the disturbance of type I is not really evolved in the environment dynamics, the weight $\omega_k$ is thus set to be zero during the whole process. Further, to extract the true information, we set the weight $\lambda_k$ adaptively according to the timely estimation of the environment. Since the estimate covariance $\Sigma_k$ is, in general, decreased as more measurements are absorbed during the filtering process, it can be seen that the sequence $\{\lambda_k\}_{k \in \mathbb{N}_+}$ will increase with an upper bound set to be $\bar{\lambda}$. 
With the help of Proposition~\ref{prop:DUCB_I}, we are now ready to present the result of regret analysis for our algorithm.

\begin{theorem}\label{thm:regretAnalysis_I}
  Suppose that $\{\mathbf{p}_{k}\}_{ k \in \mathbb{N}_+}$ is the sequence generated by Algorithm~\ref{algo:UCB} {under the conditions in Proposition~\ref{prop:DUCB_I}}, let $\bar{\lambda}$ be specified as $\bar{\lambda} = \sqrt{N}/B_K$, it holds with probability at least $1-\delta$ that,
  \begin{align}
    R_K \le \widetilde{\mathcal{O}}\Big(N^2\sqrt{K} + N^{5/2}B_K\Big),\quad \forall K >0.
    \end{align}
\end{theorem}

\begin{IEEEproof}
  See Appendix II-B.
\end{IEEEproof}

\subsection{On the Disturbance of Type II}\label{subsec:regretAnalysis_II}

Similar to the previous analysis, to provide the performance guarantee for our algorithm in this part, we also rely on the notion of regret. However, considering the difference in features of the disturbance of type II; see details in Section~\ref{subsec:SourceSeeking}, the definition of regret should be amended accordingly,
\begin{align}
	\widetilde{r}_k := \widetilde{F}_k(\mathbf{p}_k^\star) \hspace{-2pt}-\hspace{-2pt} \widetilde{F}_k(\mathbf{p}_k),
\end{align}
where the objective function $\widetilde{F}_k(\cdot)$ is defined in~\eqref{objFunc_II}. Likewise, it is also expected to show a sub-linear cumulative regret, i.e., $\widetilde{R}_K := \sum_{k=0}^K \widetilde{r}_k$ increases sub-linearly with respect to $K$.

Due to the fact that the disturbance of type II is imposed in the environment dynamics, the current state $\widetilde{\bm\phi}_k$ inherently accumulates all disturbances prior to the time $k$. As a result, it is not necessarily implied by Assumptions~\ref{assump:dynamicsBound} and \ref{assump:disturbBound} that $\widetilde{\bm\phi}_k$ is upper bounded if the sequence $B_K$ is allowed to be increased infinitely. Thus, to ensure the well-definedness of our problem, we need an additional assumption.

\begin{assumption}\label{assump:stateBound}
  There exists an uniform upper bound $\bar{\phi} > 0$ such that $\|\widetilde{\bm\phi}_k\| \le \bar{\phi}, \forall k \ge 0$.
\end{assumption}

Now, we follow the similar path as in the previous analysis to show the sub-linear regret of $\widetilde{R}_K$. Note that, due to the long term effect of the second type disturbance in the state $\widetilde{\bm\phi}_k$, one cannot expect that the D-UCB $\bm\mu_k$ serves as an upper bound for $\widetilde{\bm\phi}_k$. To deal with this issue, we construct an auxiliary variable $\overbar{\bm{\phi}}_k \in \mathbb{R}^N$, i.e.,
\begin{equation}\label{tildePhi}
  \begin{aligned}
  \overbar{\bm{\phi}}_k &:= A[k:1]\Upsilon_k^{-1}\Big(\Sigma_0^{-1} \widetilde{\bm\phi}_0 \\
  &+ \sum_{t=0}^{k-1}\lambda_tA[t:1]^\top H_t^\top V^{-1} H_t\widetilde{\bm\phi}_t + \lambda_{k-1} A[k:1]^{-1} \widetilde{\bm\phi}_{k}\Big),
\end{aligned}
\end{equation}
which helps build a connection between $\bm\mu_k$ and the state $\widetilde{\bm\phi}_k$ as shown in the following propositions.

\begin{proposition}\label{prop:DUCB_II}
	Under Assumptions~\ref{assump:dynamicsBound}--\ref{assump:stateBound} and the conditions in Proposition~\ref{prop:DUCB_I}, let $\widehat{\bm{\phi}}_k$ and $\Sigma_k$ be generated by the discounted Kalman filter~\eqref{Kalman} with $\lambda_k = \omega_k = (1/\gamma)^k$ where $0 < \gamma <1$, then it holds that, 
  \begin{align}\label{IneqUCB}
    {\mathbb{P}( \overbar{\bm\phi}_k \;{\preceq}\; \bm\mu_k ) \ge 1-\delta}, \quad \forall k \ge0,
  \end{align}
  when the sequence $\{\beta_k(\delta)\}_{k\in\mathbb{N}_+}$ in D-UCB satisfies
  \begin{equation}\label{betaK_II}
  \begin{aligned}
    &\beta_k(\delta) \ge \sqrt{N} \cdot \Bigg(C_1 + C_3 {\gamma}^{(1-k)/2} \\
  & + C_2\sqrt{N}{\gamma}^{(1-k)/2} \cdot \sqrt{\log\Big(\frac{1+\bar{\alpha}/\ubar{v}^2\cdot\sum_{t=0}^{k-1} \gamma^{2(k-t-1)}}{\delta^{2/N}}\Big)}\Bigg), 
  \end{aligned}
  \end{equation}
  where $C_1$ and $C_2$ are defined as same as in Proposition~\ref{prop:DUCB_I} and $C_3 = \bar{\phi}/\sqrt{\ubar{\alpha}}$.
\end{proposition}
\vspace{5pt}
\begin{IEEEproof}
  See Appendix II-C.
\end{IEEEproof}
\vspace{5pt}

It is proved by Proposition~\ref{prop:DUCB_II} that the D-UCB $\bm\mu_k$ provides a valid upper bound for the constructed variable $\overbar{\bm{\phi}}_k$ if $\beta_k(\delta)$ is chosen appropriately. To further build the connection between $\bm\mu_k$ and the true state $\widetilde{\bm\phi}_k$, it can be shown that the discrepancy between $\overbar{\bm{\phi}}_k$ and $\widetilde{\bm\phi}_k$ will be bounded by a term related to the disturbances $\bm\delta_k$. However, for the sake of presentation, such a result will be deferred, and we directly provide the statement of sub-linear regret for our algorithm in the following theorem.  The bound of $\|\overbar{\bm{\phi}}_k - \widetilde{\bm\phi}_k\|$ will be shown as an intermediate step of proofs for the theorem in Appendix.


\begin{theorem}\label{thm:regretAnalysis_II}
 Suppose that $\{\mathbf{p}_{k}\}_{ k \in \mathbb{N}_+}$ is the sequence generated by Algorithm~\ref{algo:UCB} {under the conditions in Proposition~\ref{prop:DUCB_II}}, let ${\gamma}$ be specified as $\gamma = 1 - (B_K/K)^{2/3}$, then it holds that with probability $1-\delta$,
  \begin{align}
    \widetilde{R}_K \le \widetilde{\mathcal{O}}\Big( N^2B_K^{1/3}K^{2/3}\Big),\quad \forall K >0.
    \end{align}
\end{theorem}
\begin{IEEEproof}
  See Appendix II-D
\end{IEEEproof}


\begin{remark}
  Note that, in Proposition~\ref{prop:DUCB_II} and Theorem~\ref{thm:regretAnalysis_II}, the two weights $\lambda_k$ and $\omega_k$ are specified as $\lambda_k = \omega_k = (1/\gamma)^k$ where $\gamma < 1$. This means that they will increase exponentially with respect to the time-step $k$. Therefore, numerical overflow may arise in the discounted Kalman filtering as shown in~\eqref{Kalman}, when $k$ is large. To deal with this issue, we notice that the discounted Kalman filter, when $\lambda_k$ and $\omega_k$ are chosen as above, can be implemented equivalently by the following recursions,
  \begin{subequations}\label{implementation}
  \begin{align}
    &\widetilde{\Sigma}_{k+\shalf} = \big(\gamma\widetilde{\Sigma}_{k}^{-1} + Y_k \big)^{-1},\\
    &\widehat{\bm{\phi}}_{k+\shalf} = \widehat{\bm{\phi}}_{k} + \widetilde{\Sigma}_{k+\scriptscriptstyle{1/2}}(\mathbf{y}_k - Y_k\widehat{\bm{\phi}}_{k}),\\
    &\widetilde{\Sigma}_{k+1} = A_{k+1}\Big(\widetilde{\Sigma}_{k+\shalf}^{-1}+(1 -\gamma)\Gamma_k^{-1}\Big)^{-1}A_{k+1}^\top,\\
    &\widehat{\bm{\phi}}_{k+1} =\Big(A_{k+1} - (1 - \gamma)\widetilde{\Sigma}_{k+1}A_{k+1}^{-\top}\Gamma_k^{-1}\Big)\widehat{\bm{\phi}}_{k+\shalf},
  \end{align}
  where $\Gamma_k$ is defined as same as before. It should be  also noted that, in~\eqref{implementation}, the covariance is slightly different from the one in~\eqref{Kalman}, in the sense that $\widetilde{\Sigma}_{k} = (1/\gamma)^{k-1} \Sigma_k$. This needs to be taken into account in Algorithm~\ref{algo:UCB} when generating the D-UCB $\bm\mu_k$ by using $\widetilde{\Sigma}_{k}$.
 \end{subequations} 
\end{remark}

\subsection{Further Discussions}

Before the end of this section, a few more remarks should be added on the obtained results of the above regret analysis.

First, to tackle with the two types of non-stochastic disturbances, it can be seen from the propositions that the sequences of weights $\lambda_k$ and $\omega_k$ are also determined differently. More specifically, for the external disturbance which only affects the measured state $\widetilde{\bm\phi}_k$ but not evolve with the nominal dynamics, the sequence of $\lambda_k$ is chosen as increased at the same rate of $1/\|Y_k\|_{\Sigma_k}$. This is due to the fact that the disturbance $\bm\delta_k$ in this case only comes into play when the state is measured, and thus the weight $\lambda_k$ is also adjusted according to the measurement information in $Y_k$ and the current progress on $\Sigma_k$. Since the covariance $\Sigma_k$, which basically suggests the uncertainty of~our estimation, decreases as more measurements are absorbed in the estimation, the weight $\lambda_k$ is increased during the process, meaning that the measurement received later are more trusted. For the internal disturbance, the sequence of $\lambda_k$ is also chosen as increased, but at a fixed exponential rate of $(1/\gamma)^k$. Another primary difference is that, while $\omega_k$ is set to be zero previously, here we let $\omega_k$ increase at the same exponential rate of $(1/\gamma)^k$. The reason for such a difference can be explained as follows. Since the internal disturbance, regardless of the measurements, is accumulated during the whole process, an additional weight needs to be incorporated to deal with it globally, and therefore the increasing $\omega_k$ is introduced to decrease the covariance $\Sigma_k$ accordingly. 
Note that this does not mean the uncertainty of our estimation is decreased brutally, as in the D-UCB $\bm\mu_k$, the sequence of $\beta_k(\delta)$ is also increased by an extra term related to $1/\gamma$ to adjust our construction of the confidence bound.

Second, it can be concluded by the two theorems that once the disturbance bound $B_K$ increases sub-linearly, the regrets generated by our algorithm for both cases also grow at a~sub-linear rate, meaning that the agents will be able to track the moving sources dynamically under the disturbed environment. More precisely, while the regret for the first case increases at the rate of $\widetilde{\mathcal{O}}(\sqrt{K} + B_K)$, the rate is $\widetilde{\mathcal{O}}(B_K^{1/3}K^{2/3})$ for the second case. Note that both of them are identical to the state-of-the-art results in the study of bandit algorithms~with non-stationary and adversarial settings. Therefore, we can conclude that our developments of discounted Kalman filter and D-UCB do not degrade the performance of algorithm with respect to its convergence. However, in terms of the scale of~the problem~$N$, i.e., size of the searching environment, the complexity of our algorithm indeed grows at the rate of $\widetilde{\mathcal{O}}(N^2)$, as compared to $\widetilde{\mathcal{O}}(N)$ in the literature. This is mainly due to the reason~that the ellipsoid confidence sets in the classical UCB-based methods is changed to the polytope one in our algorithm. Despite this fact, we argue that such an increase of complexity is actually reasonable, since more computational complexities have been reduced by avoiding the combinatorial problems at each step.

\section{Simulation}

In this section, numerical examples are provided to validate the effectiveness of our multi-agent source seeking algorithm. We consider a pollution monitoring application where three mobile robots are deployed in a pollution diffusion field with the aim to localize as many leaking sources as possible. The dynamics of the pollution field is governed a convection diffusion equation.
More details of the simulation settings can be found in~\cite{du2021multi}, including linearization of the partial differential equation, robots' measurement models and their communication topology, specification of the pollution field, etc. However, a key difference here is that the non-stochastic disturbances are assumed to be present after the linearization of the dynamics. More concretely, the linearized model of the pollution field is represented by 
\begin{align}
  \bm{\Phi}_{k+1} = A \mathbf{\bm{\Phi}}_k + \bm{\delta_k},
\end{align}
where $\mathbf{\bm{\Phi}}_k$ denotes the discretized states of the field, $A$ is the state transition matrix and $\bm{\delta_k}$ represents the non-stochastic disturbance.

In particular, we consider in this simulation that the pollution field is modeled by a $D \times D$ lattice with $D = 50$.~Each of the mobile robots is capable of sensing a circular area with radius $R = 5$ during the searching process. The sensing noise is assumed to be \textit{i.i.d.} Gaussian with zero-mean and covariance $V^i = 4\cdot\mathbf{I}_m, i=1,2,3$. In terms of the disturbance, we here consider two different scenarios: i) a slowly-varying disturbance which occurs externally; and ii) an abruptly-changing which occurs internally. For the slowly-varying disturbance of type I, it is assumed that $\bm{\delta}_k = 0$ when $k < 100$ and $\bm{\delta}_k = 1/k^2\cdot \bm{\Pi}_0$ when $k \ge 100$ where $\bm{\Pi}_0$ is randomly generated. For the abruptly-changing disturbance of type II, we consider that two more leaking sources are randomly injected into the field during the period of $[150, 165]$ and $[600, 615]$. That is, $\bm{\delta}_k = \bm{\Pi}_1$ for $150 \le k\le 165$ and $\bm{\delta}_k = \bm{\Pi}_2$ for $600 \le k\le 615$ where $\bm{\Pi}_1, \bm{\Pi}_2 \in \mathbb{R}^N$ are randomly generated, and $\bm{\delta}_k = 0$ otherwise. 

\begin{figure}
  \begin{subfigure}{0.475\linewidth}
    \centering
\includegraphics[width=\textwidth]{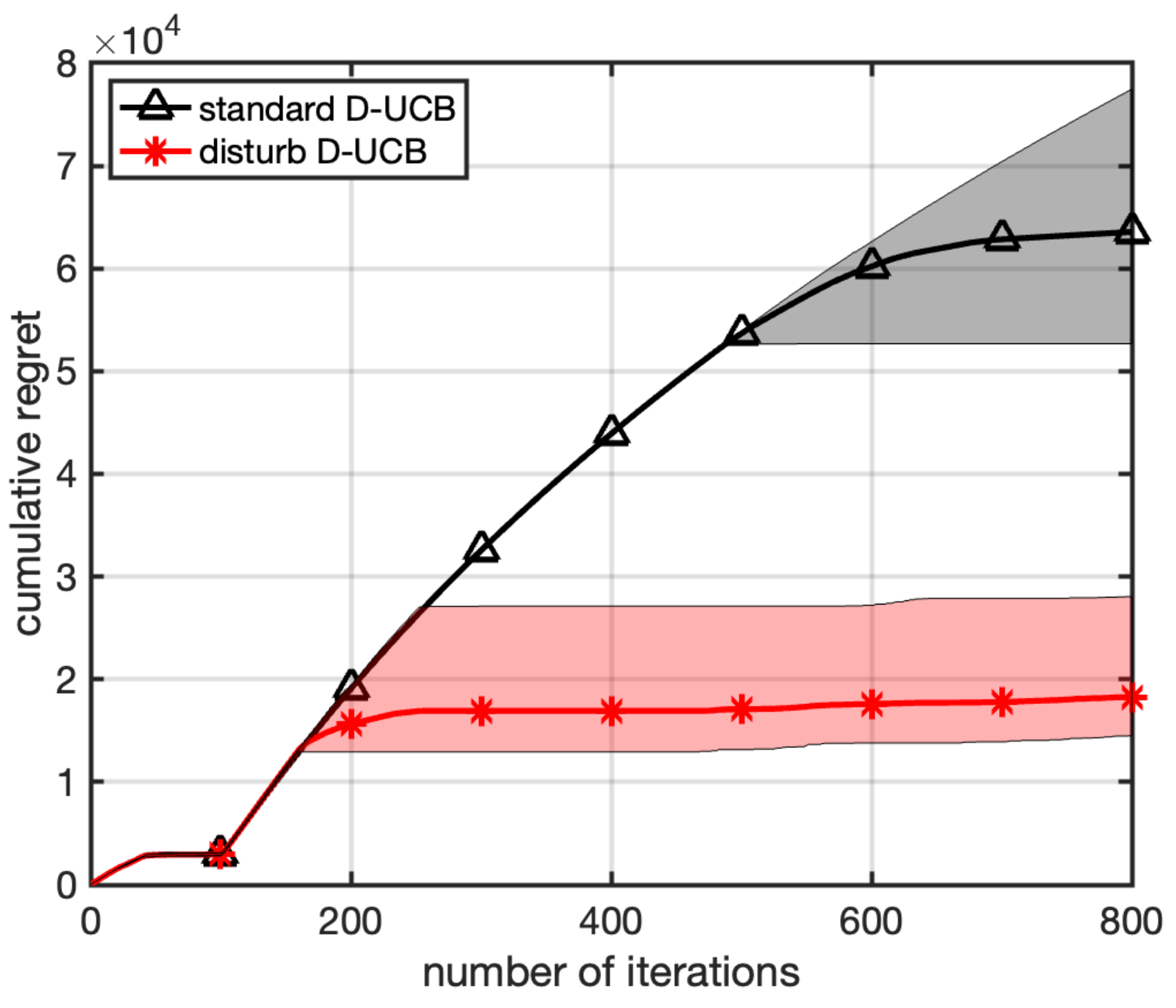}
    \label{fig:2}
    \vspace{-12pt}
    \caption{Slowly-varying disturbance}
  \end{subfigure}
    \begin{subfigure}{0.49\linewidth}
  \centering
\includegraphics[width=\textwidth]{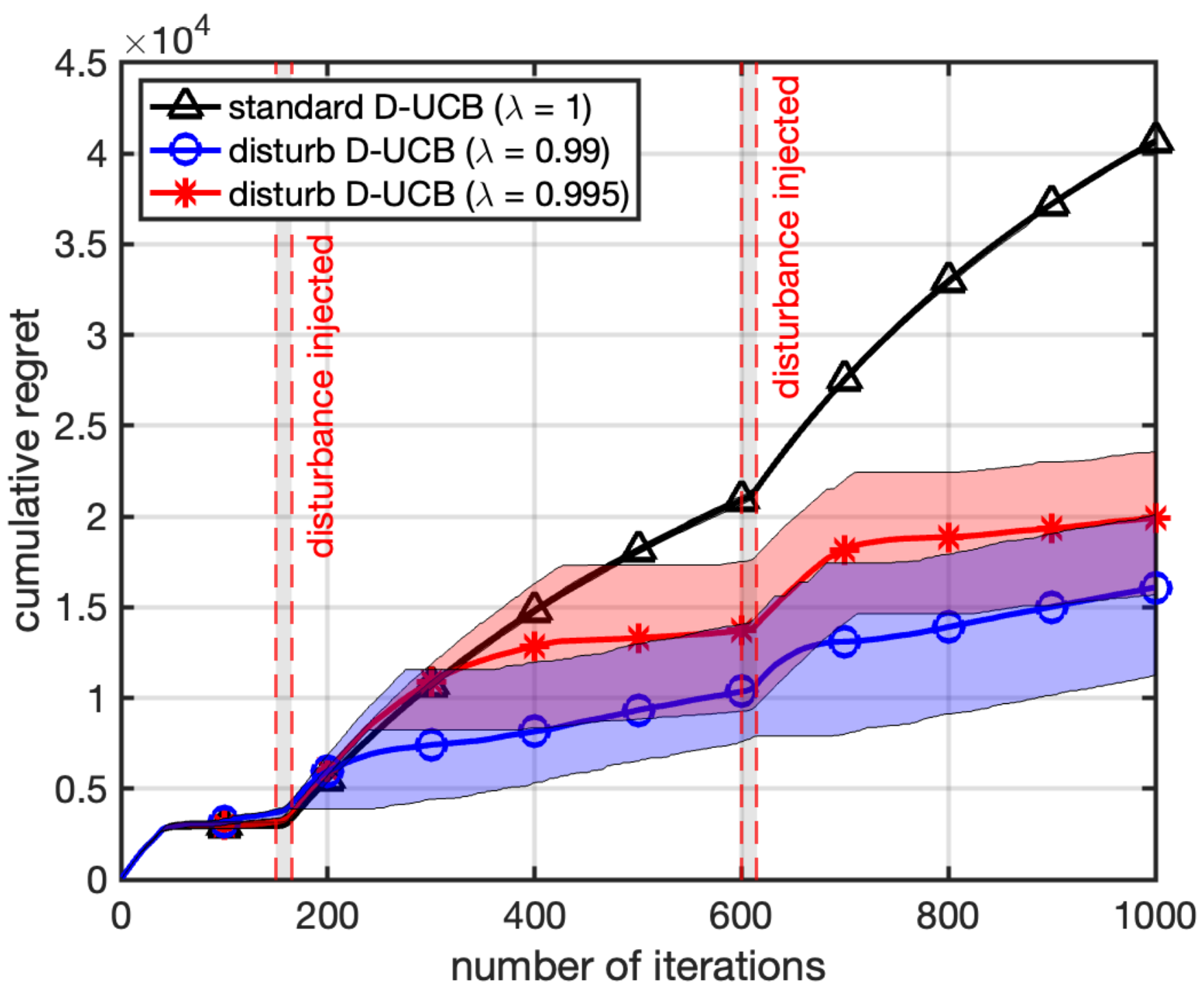}
\caption{Abruptly-changing disturbance}
\label{subfig:2}
 \end{subfigure}
\centering
  \caption{Regrets generated by the algorithms under different scenarios}
  \label{fig:regret}
\end{figure}

To illustrate the performance of our algorithm in seeking the dynamical pollution sources with the two types of disturbances, we show the cumulative~regrets~$R_T$ produced by Algorithm~\ref{algo:UCB}, respectively. The obtained numerical results are shown in Fig.~\ref{fig:regret}, in which each curve is corresponding to $20$ independent trials. It can be observed~from the figures that our algorithm produces the smaller cumulative regret than the one generated by the standard~\mbox{D-UCB} algorithm.  We can thus conclude that, while the standard D-UCB algorithm fails to localize the sources when the disturbances are present in the field, our algorithm manages to complete the task in both scenarios with the external and internal disturbances. More specifically, with respect to the internal abruptly-changing disturbance, we also compare the performance of our algorithm with different choices of the parameter $\gamma$. Note that by setting $\gamma = 1$, our algorithm will be naturally reduced to standard D-UCB algorithm. It can be observed that, after the disturbance are injected, our algorithm will soon adapt to the disturbed pollution field and then track the newly-added sources accordingly. On the contrary, the standard D-UCB algorithm fails to do so. In addition, it can be also seen from Fig.~\ref{subfig:2} that the smaller $\gamma$ results in a shorter period of the adaption process. This is mainly due to the fact that the agents tend to perform more explorations when the small $\gamma$ is chosen. As a result of the classical dilemma between exploration and exploitation, however, an disadvantage of the smaller $\gamma$ is that the cumulative regret grows more rapidly after the sources are localized.



\section{Conclusion}
In this paper, a learning based algorithm is developed to solve the problem of multi-agent online source seeking under the environment disturbed by non-stochastic perturbations. Building on the technique of discounted Kalman filtering as well as the notion of D-UCB proposed in our previous work, our algorithm enables the computation-efficient cooperation among the multi-agent network and is robust against the non-stochastic perturbations (also interpreted as the adversarial disturbances in the context of multi-armed bandits). It is shown that a sub-linear cumulative regret is achieved by our algorithm, which is comparable to the state-of-art. Numerical results on a real-world pollution monitoring application is finally provided to support our theoretical findings.

\section*{Appendix I: proof of Lemma~\ref{lemma:estRecursion}}\label{sec:appendix_I}

 Let us prove Lemma~\ref{lemma:estRecursion} by mathematical induction. First, it is straightforward to confirm that, given the initialization $\widehat{\bm\phi}_0$, $\Sigma_0$, and $\omega_{-1}= 0 $, the recursions~\eqref{Kalman} and \eqref{Kalman_recursion} produce the identical $\widehat{\bm\phi}_1$ and $\Sigma_1$. Next, we assume that \eqref{Kalman_recursion} generates the same results as~\eqref{Kalman} up to the time-step $k$, it will suffice to prove the consistency for the time-step $k+1$.


  In fact, based on the recursion of $\Sigma_k$ in~\eqref{Kalman}, we can have
  \begin{equation}\label{proof_Sigma}
  \begin{aligned}
    &\Sigma_{k+1} = A_{k+1}\Big(\Sigma_{k}^{-1} + \lambda_kY_k + (\omega_k - \omega_{k-1}) \Gamma_k^{-1} \Big)^{-1}A_{k+1}^\top\\
    &=A_{k+1}\Big(A[k:1]^{-\top}\Upsilon_kA[k:1]^{-1} + \lambda_kY_k \\
    & \hspace{40pt}+ (\omega_k - \omega_{k-1}) \Gamma_k^{-1} \Big)^{-1}A_{k+1}^\top\\
    &=A[k\hspace{-2pt}+\hspace{-2pt}1:1] \Big(\Upsilon_k + A[k:1]^{\top}Y_kA[k:1]\\
    &\hspace{40pt}  + (\omega_k - \omega_{k-1})\mathbf{I}_N\Big)^{-1}A[k\hspace{-2pt}+\hspace{-2pt}1:1]^\top\\
    &=A[k\hspace{-2pt}+\hspace{-2pt}1:1]\Upsilon_{k+1}^{-1}A[k\hspace{-2pt}+\hspace{-2pt}1:1]^{\top},
  \end{aligned}
  \end{equation}
  where the second equality comes from our assumption of $\Sigma_k$ in the form of~\eqref{Kalman_recursion_Sigma} and the last equality is due to the definition of $\Upsilon_k$ in~\eqref{upsilon}.
  Similarly, based on the recursion of $\widehat{\bm\phi}_k$ in~\eqref{Kalman}, we can have
  \begin{equation}\label{proof_phi}
  \begin{aligned}
    &\widehat{\bm{\phi}}_{k+1} = A[k\hspace{-2pt}+\hspace{-2pt}1:1] \Upsilon_{k+1}^{-1}A[k\hspace{-2pt}+\hspace{-2pt}1:1]^\top A_{k+1}^{-\top} \Sigma_{k+\shalf}^{-1}\widehat{\bm{\phi}}_{k+\shalf}\\
    &=A[k\hspace{-2pt}+\hspace{-2pt}1:1] \Upsilon_{k+1}^{-1}A[k:1]^\top \Big( \Sigma_{k+\shalf}^{-1}\widehat{\bm{\phi}}_{k} + \lambda_k(\mathbf{y}_k - Y_k\widehat{\bm{\phi}}_{k}) \Big)\\
    &= A[k\hspace{-2pt}+\hspace{-2pt}1:1] \Upsilon_{k+1}^{-1}A[k:1]^\top \Big(\Sigma_{k}^{-1}\widehat{\bm{\phi}}_{k} +\lambda_k\mathbf{y}_k\Big)\\
    & = A[k\hspace{-2pt}+\hspace{-2pt}1:1] \Upsilon_{k+1}^{-1}\Big(\Upsilon_k A[k:1]^{-1}\widehat{\bm{\phi}}_{k} + \lambda_kA[k:1]^\top\mathbf{y}_k\Big)\\
    & = A[k\hspace{-2pt}+\hspace{-2pt}1:1] \Upsilon_{k+1}^{-1}\Big(\Sigma_0^{-1} \widehat{\bm{\phi}}_0 + \sum_{t=0}^{k}\lambda_tA[t:1]^\top \mathbf{y}_t\Big),
  \end{aligned}
  \end{equation}
  where the first equality comes from~\eqref{proof_Sigma} which just has been proved; the second and third equalities are due to~\eqref{Kalman}; the second last equality follows $\Sigma_k$ in the form of~\eqref{Kalman_recursion_Sigma}; and the last one is due to our assumption of $\widehat{\bm{\phi}}_{k}$ in the form of~\eqref{Kalman_recursion_phi}.

\section*{Appendix II: proofs of Main Theorems}\label{sec:appendix_I}
We shall notice that the proofs in this section are mainly inspired by~\cite{russac2019weighted} and~\cite{he2022nearly}, which performed the regret analysis in the context of stochastic linear bandits under non-stationary and adversarial environments, respectively. The contributions of our proofs are i) integration of linear dynamics and Kalman filtering into the algorithmic framework; and ii) adaptation of the new notion of D-UCB into the regret analysis.

To facilitate the following proofs, let us start by introducing some useful vector norms. First, associated with the diagonal matrix of an arbitrary positive definite matrix $M\in \mathbb{R}^{N \times N}$, i.e., $\mathcal{D}_M = \text{Diag}\{m_{11}, m_{22}, \cdots, m_{NN}\} \in \mathbb{R}^{N \times N}$, we define the $\mathcal{L}_2$-based vector norm $\|\cdot\|_{\mathcal{D}_M} : \mathbb{R}^N \to \mathbb{R}_+$ as 
\begin{align}\label{2norm}
  \|\mathbf{x}\|_{\mathcal{D}_M} := \sqrt{\sum_{i=1}^N m_{ii}\cdot x_i^2},
\end{align}
where $\mathbf{x} = [x_1, x_2, \cdots x_N]^\top \in \mathbb{R}^N$. Further, let us define~the $\mathcal{L}_\infty$-based norm $\|\cdot\|_{\mathcal{D}_M, \infty}: \mathbb{R}^N \to \mathbb{R}_+$ with respect to the matrix $\mathcal{D}_M$ as
\begin{align}\label{Linfty}
  \|\mathbf{x}\|_{\mathcal{D}_M, \infty} := \max_{1 \le i\le N} \; m_{ii}\cdot|x_i|.
\end{align}
Note that the above norm $\|\cdot\|_{\mathcal{D}_M, \infty}$ is well-defined since the positive definiteness of $M$ ensures that $m_{ii} > 0$. Similarly, we define the $\mathcal{L}_1$-based norm \mbox{$\|\cdot\|_{\mathcal{D}_M, 1}: \mathbb{R}^N \to \mathbb{R}_+$} as
\begin{align}\label{L1}
  \|\mathbf{x}\|_{\mathcal{D}_M,1} := \sum_{i=1}^N m_{ii}\cdot|x_i|.
\end{align}

With the vector norms introduced above, it can be immediately verified that $\|\cdot\|_{\mathcal{D}_M, 1}$ and $\|\cdot\|_{\mathcal{D}_M^{-1}, \infty}$ are dual norms where $\mathcal{D}_M^{-1}$ takes the inverse of the matrix $\mathcal{D}_M$. In addition, we provide in the following lemma the connections among all defined norms.

 \begin{lemma}\label{lemma:diagNorm}
 For arbitrary positive definite matrix $M$, it holds that, 1) $\|\mathbf{x}\|_{\mathcal{D}_M,\infty}\le \|\mathbf{x}\|_{\mathcal{D}_M^{2}}$; 2) $\|\mathbf{x}\|_{\mathcal{D}_M,1} \le \sqrt{N}\cdot \|\mathbf{x}\|_{\mathcal{D}_M^{2}}$; and 3) $\|\mathbf{x}\|_M \le {\sqrt{N} \cdot \|\mathbf{x}\|_{\mathcal{D}_M}}$.
\end{lemma}

\begin{IEEEproof}
	While the inequalities a) and b) can be straightforwardly confirmed by the definitions and the inequality of arithmetic and geometric means, respectively, the part c) is proved as follows.
	  \begin{equation}
    \begin{aligned}
      \|\mathbf{x}\|_M^2 
       & \le \sum_{i=1}^N m_{ii}\cdot x_i^2 + \sum_{i=1}^N \sum_{j\neq i} |m_{ij}|\cdot|x_i x_j|\\
      & \le \sum_{i=1}^N m_{ii}\cdot x_i^2 + \sum_{i=1}^N \sum_{j\neq i} \sqrt{m_{ii}m_{jj}}\cdot|x_i x_j|\\
      & \le \sum_{i=1}^N m_{ii}\cdot x_i^2 + \sum_{i=1}^N \sum_{j\neq i} \frac{1}{2}(m_{ii}\cdot x_i^2 + m_{jj}\cdot x_j^2)\\
      & = N \cdot \|\mathbf{x}\|^2_{\mathcal{D}_M}.
    \end{aligned}
  \end{equation}
  Note that the first inequality is due to the positive definiteness of $M$, i.e., $|m_{ij}| \le \sqrt{m_{ii}m_{jj}}$. Hence, the proof is completed.
\end{IEEEproof}

\setcounter{subsection}{0}

\subsection{Proof of Proposition~\ref{prop:DUCB_I}}

To prove the inequality~\eqref{IneqUCB_I} in Proposition~\ref{prop:DUCB_I}, with the help of the above defined vector norms, it will suffice to show
\begin{align}\label{IneqUCBinftyNorm}
  \mathbb{P}\Big(\big\|\widehat{\bm{\phi}}_k - \bm{\phi}_k\big\|_{\mathcal{D}^{-1/2}_{\Sigma_k},\infty} \le \beta_k(\delta)\Big) \ge 1-\delta.
\end{align}
Note that the inequality in~\eqref{IneqUCBinftyNorm} is stronger than the one in~\eqref{IneqUCB_I}, in the sense that the state $\bm{\phi}_k$ is both upper and lower bounded. Though the lower bound is not reflected in the development of our algorithm, it helps the proof for the sub-linear regret. In addition, due to the fact that the weight $\omega$ is specified as $\omega_k \equiv 0$ in this part, it indeed changes the generation of state estimates by simplifying the matrix $\Upsilon_k$ as
\begin{align}\label{upsilon_change}
	\Upsilon_k := \Sigma_0^{-1} + \sum_{t=0}^{k-1}\lambda_tA[t:1]^\top Y_t A[t:1].
\end{align}

According to the nature of the first type of disturbance, the disturbed state can be expressed as $\widetilde{\bm\phi}_k = A[k:1] \bm{\phi}_0 + \bm{\delta}_k$ and thus the measurement is $\mathbf{z}_k =H_k (A[k:1] \bm{\phi}_0+\bm\delta_k) + \mathbf{n}_k$. Then, by Lemma~\ref{lemma:estRecursion} and the definitions of $\Upsilon_k$ and $Y_k$, the state estimate $\widehat{\bm{\phi}}_k$ has
\begin{equation}
  \begin{aligned}
    \widehat{\bm{\phi}}_k 
    &= A[k:1]\Upsilon_k ^{-1}\Big(\Sigma_0^{-1} \widehat{\bm{\phi}}_0 +\sum_{t=0}^{k-1} \lambda_tA[t:1]^\top H_t^\top V^{-1}\mathbf{n}_t \\
    &\hspace{10pt}+\sum_{t=0}^{k-1} \lambda_tA[t:1]^\top Y_t \big(A[t:1] {\bm{\phi}}_0 +\bm{\delta}_t\big)\Big)\\ 
      & =\bm{\phi}_k +A[k:1]\Upsilon_k^{-1}\Big(\sum_{t=0}^{k-1} \lambda_t A[t:1]^\top H_t^\top V^{-1} \mathbf{n}_t \\
      &\hspace{10pt}+\sum_{t=0}^{k-1} \lambda_tA[t:1]^\top Y_t \bm{\delta}_t + \Sigma_0^{-1} ( \widehat{\bm{\phi}}_0 - {\bm{\phi}}_0) \Big).
  \end{aligned}
\end{equation}
Therefore, it holds for $\forall \mathbf{x} \in \mathbb{R}^{N}$ that
\begin{equation}\label{errorMeasure}
  \begin{aligned}
    &\mathbf{x}^\top(\widehat{\bm{\phi}}_k - \bm{\phi}_k)\\
      &\hspace{-5pt}\overset{(1.a)}{\le} \big\|A[k:1]^\top \mathbf{x}\big\|_{\Upsilon_k^{-1}}\cdot \Bigg(\big\|\Sigma_0^{-1} ( \widehat{\bm{\phi}}_0 - {\bm{\phi}}_0)\big\|_{\Upsilon_k^{-1}} \\
      &\hspace{10pt}+ \Big\|\sum_{t=0}^{k-1} \lambda_tA[t:1]^\top Y_t \bm{\delta}_t\Big\|_{\Upsilon_k^{-1}} \\
      &\hspace{10pt} + \Big\|\sum_{t=0}^{k-1}\lambda_tA[t:1]^\top H_t^\top V^{-1} \mathbf{n}_t\Big\|_{\Upsilon_k^{-1}}\Bigg)\\
      & \hspace{-5pt}\overset{(1.b)}{=} \big\|\mathbf{x}\big\|_{\Sigma_k} \cdot\Bigg( \big\|\Sigma_0^{-1} ( \widehat{\bm{\phi}}_0 - {\bm{\phi}}_0)\big\|_{\Upsilon_k^{-1}} \\
      &\hspace{10pt} + \Big\|\sum_{t=0}^{k-1}\lambda_tA[t:1]^\top H_t^\top V^{-1} \mathbf{n}_t\Big\|_{\Upsilon_k^{-1}}\\
      &\hspace{10pt}+ \Big\|\sum_{t=0}^{k-1} \lambda_tA[t:1]^\top Y_t \bm{\delta}_t\Big\|_{\Upsilon_k^{-1}}\Bigg)\\
	&\hspace{-5pt}\overset{(1.c)}{\le} {\sqrt{N}} \cdot \big\|\mathbf{x}\big\|_{\mathcal{D}_{\Sigma_k}} \cdot\Bigg( \big\|\Sigma_0^{-1} ( \widehat{\bm{\phi}}_0 - {\bm{\phi}}_0)\big\|_{\Upsilon_k^{-1}}\\
      &\hspace{10pt}+ \Big\|\sum_{t=0}^{k-1} \lambda_tA[t:1]^\top Y_t \bm{\delta}_t\Big\|_{\Upsilon_k^{-1}} \\
      &\hspace{10pt} + \Big\|\sum_{t=0}^{k-1}\lambda_tA[t:1]^\top H_t^\top V^{-1} \mathbf{n}_t\Big\|_{\Upsilon_k^{-1}}\Bigg),
	\end{aligned}
\end{equation}
where $(1.a)$ is according to the Cauchy-Schwartz and triangle inequalities; $(1.b)$ is due to the recursion of $\Sigma_k$ in the form of~\eqref{Kalman_recursion_Sigma}; and $(1.c)$ is based on Lemma~\ref{lemma:diagNorm}-3).

Now, by Lemma~\ref{lemma:diagNorm}-1), it follows that
\begin{equation}\label{keyIneq}
  \begin{aligned}
    \big\|\widehat{\bm{\phi}}_k &- \bm{\phi}_k\big\|_{\mathcal{D}^{-1/2}_{\Sigma_k}, \infty} \le \big\|\widehat{\bm{\phi}}_k - \bm{\phi}_k\big\|_{\mathcal{D}^{-1}_{\Sigma_k}} \\
    &\le  {\sqrt{N}} \cdot\Bigg(\big\|\Sigma_0^{-1} ( \widehat{\bm{\phi}}_0 - {\bm{\phi}}_0)\big\|_{\Upsilon_k^{-1}} \\
      &\hspace{10pt}+ \Big\|\sum_{t=0}^{k-1} \lambda_tA[t:1]^\top Y_t \bm{\delta}_t\Big\|_{\Upsilon_k^{-1}} \\
      & \hspace{10pt}+\Big\|\sum_{t=0}^{k-1}\lambda_tA[t:1]^\top H_t^\top V^{-1} \mathbf{n}_t\Big\|_{\Upsilon_k^{-1}} \Bigg),
  \end{aligned}
\end{equation}
where the last inequality is according to~\eqref{errorMeasure} and meanwhile taking $\mathbf{x} = \mathcal{D}^{-1}_{\Sigma_k}(\widehat{\bm{\phi}}_k - \bm{\phi}_k)$. Next, to prove the inequality~\eqref{IneqUCBinftyNorm}, we upper bound the three terms on the right hand side of~\eqref{keyIneq} in the following three lemmas, respectively.

\begin{lemma}\label{lemma:upperBoundI}
  Under the conditions in Proposition~\ref{prop:DUCB_I}, there exists a constant $C_1 = \|\widehat{\bm{\phi}}_0 - \bm{\phi}_0\| /\sqrt{\ubar{\sigma}}$ such that,
  \begin{align}
    \big\|\Sigma_0^{-1} ( \widehat{\bm{\phi}}_0 - {\bm{\phi}}_0)\big\|_{\Upsilon_k^{-1}} \le C_1, \quad \forall k \ge 0.
  \end{align}
\end{lemma}

\begin{IEEEproof}
  By the definition of the matrix $\Upsilon_k$, it is straightforward to see that ${\Upsilon_k^{-1}} \;{\preceq}\; \Sigma_0$, and therefore,
  \begin{equation}
      \begin{aligned}
    &\big\|\Sigma_0^{-1} ( \widehat{\bm{\phi}}_0 - {\bm{\phi}}_0)\big\|^2_{\Upsilon_k^{-1}}\\ &=  ( \widehat{\bm{\phi}}_0 - {\bm{\phi}}_0)^\top\Sigma_0^{-1}\Upsilon_k^{-1}\Sigma_0^{-1} ( \widehat{\bm{\phi}}_0 - {\bm{\phi}}_0)\\
    & \le ( \widehat{\bm{\phi}}_0 - {\bm{\phi}}_0)^\top\Sigma_0^{-1} ( \widehat{\bm{\phi}}_0 - {\bm{\phi}}_0)\\
    & \le 1/\ubar{\sigma} \cdot \|\widehat{\bm{\phi}}_0 - \bm{\phi}_0\|^2,
  \end{aligned}
  \end{equation}
  where the last inequality is due to the assumption $\Sigma_0 \succeq \ubar{\sigma} \cdot \mathbf{I}$. Thus, the proof is completed.
\end{IEEEproof}

\begin{lemma}\label{lemma:upperBoundII}
  Under the conditions in Proposition~\ref{prop:DUCB_I}, one can have that
  \begin{align}
  \Big\|\sum_{t=0}^{k-1} \lambda_tA[t:1]^\top Y_t \bm{\delta}_t\Big\|_{\Upsilon_k^{-1}}\le \bar{\lambda}B_{k},
  \end{align}
  where the sequence $\{B_k\}_{k \in \mathbb{N}_+}$ is defined in Assumption~\ref{assump:disturbBound}.
\end{lemma}
\begin{IEEEproof}
	It holds that
	\begin{equation}
		\begin{aligned}
			&\Big\|\sum_{t=0}^{k-1} \lambda_tA[t:1]^\top Y_t \bm{\delta}_t\Big\|_{\Upsilon_k^{-1}}\overset{(2.a)}{\le} \sum_{t=0}^{k-1} \lambda_t \big\|A[t:1]^\top Y_t\big\|_{\Upsilon_t^{-1}} \| \bm{\delta}_t\|\\
			&\hspace{-5pt}\overset{(2.b)}{=} \sum_{t=0}^{k-1} \lambda_t \big\| Y_t\big\|_{\Sigma_t}\| \bm{\delta}_t\|\overset{(2.c)}{\le} \sum_{t=0}^{k-1}\| \bar{\lambda}\bm{\delta}_t\|\overset{(2.d)}{\le} \bar{\lambda}B_{k},
		\end{aligned}
	\end{equation}
	where $(2.a)$ is due to the triangle inequality and the fact that $\Upsilon_k \ge \Upsilon_t, \forall k \ge t$ by~\eqref{upsilon_change}; $(2.b)$ is based on the recursion~\eqref{Kalman_recursion_Sigma} of $\Sigma_t$; $(2.c)$ is due to the specification of the weight $\lambda_k$, i.e., $\lambda_k = \min\{1, \bar{\lambda}/\|Y_t\|_{\Sigma_t}\}$; and $(2.d)$ is by Assumption~\ref{assump:disturbBound}.
\end{IEEEproof}

\begin{lemma}\label{lemma:upperBoundIII}
  Under the conditions in Proposition~\ref{prop:DUCB_I}, there exists a constant $C_2 = \bar{v}^2 \sqrt{\max\{2,2/\ubar{v}\}}$ such that the following inequality holds with probability at least $1-\delta$,
  \begin{equation}
  \begin{aligned}
    &\Big\|\sum_{t=0}^{k-1}\lambda_tA[t:1]^\top H_t^\top V^{-1} \mathbf{n}_t\Big\|_{\Upsilon_k^{-1}} \\
    &\hspace{50pt}\le C_2\sqrt{N} \cdot \sqrt{\log\Big(\frac{\bar{\sigma}/\ubar{\sigma} + \bar{\alpha}\bar{\sigma}\cdot k / \ubar{v}^2}{\delta^{2/N}}\Big)}.
  \end{aligned}
  \end{equation}
 \end{lemma}

 \begin{IEEEproof}
    This proof is based on the existing results on the self-normalized Martingale, see e.g.,~\cite{russac2019weighted}. For the notational simplicity, let us define
    \begin{align}\label{X}
      X_t := \lambda_tA[t:1]^\top H_t^\top V^{-1} \in \mathbb{R}^{N \times M}.
    \end{align}
    Then, according to the result of self-normalized Martingale, it holds with probability at least $1-\delta$ that,
    \begin{align}\label{upperBoundI}
      \Big\|\sum_{t=0}^{k-1} X_t\mathbf{n}_t\Big\|_{\Omega_k^{-1}} \le 2\bar{v}^2 \cdot\sqrt{\log\Big(\frac{\det(\Omega_k)^{1/2}\det(\Sigma_0)^{1/2}}{\delta}\Big)},
    \end{align}
    where $\Omega_k := \Sigma_0^{-1} + \sum_{t=0}^{k-1} X_t X_t^\top\in\mathbb{R}^{N \times N}$. Note that there is a slight difference between $\Omega_k$ and $\Upsilon_k$, and we show there exists a constant $C'_2  = \max\{1, 1/\ubar{v}\}$ such that $\Omega_k \;{\preceq}\; C'_2 \Upsilon_k$. In fact, it holds that
    \begin{equation}\label{Omega_ineq}
    \begin{aligned}
      \Omega_k &= \Sigma_0^{-1} +  \sum_{t=0}^{k-1}\lambda_t^2A[t:1]^\top H_t^\top V^{-2} H_t A[t:1] \\
      &\;{\preceq}\; \Sigma_0^{-1} + 1/\ubar{v}\cdot \sum_{t=0}^{k-1}\lambda_tA[t:1]^\top H_t^\top V^{-1} H_t A[t:1] \\
      &\;{\preceq}\; \max\{1, 1/\ubar{v}\} \cdot \Upsilon_k.
    \end{aligned}
    \end{equation}
    Note that the first inequality is due to $\omega_t \le 1$ and the assumption $\ubar{v}\cdot\mathbf{I}_M\preceq V\preceq\bar{v}\cdot\mathbf{I}_M$. Therefore, the previous statement can be immediately verified by letting $C'_2 = \max\{1, 1/\ubar{v}\}$ and implies that $\Upsilon_k^{-1} {\preceq}\; C'_2 \Omega_k^{-1}$. Together with the inequality \eqref{upperBoundI}, it follows that
    \begin{equation}
      \begin{aligned}
         &\Big\|\sum_{t=0}^{k-1} X_t\mathbf{n}_t\Big\|_{\Upsilon_k^{-1}} \le \sqrt{C'_2} \cdot \Big\|\sum_{t=0}^{k-1} X_t\mathbf{n}_t\Big\|_{\Omega_k^{-1}}\\
         &\le 2\bar{v}^2\sqrt{\max\{1, 1/\ubar{v}\}} \cdot\hspace{-2pt}\sqrt{\log\Big(\frac{\det(\Omega_k)^{1/2}\det(\Sigma_0)^{1/2}}{\delta}\Big)}.
      \end{aligned}
    \end{equation}

   Moreover, based on the inequality of arithmetic and geometric means and the definition of $\Omega_k$, it holds that
    \begin{equation}
      \begin{aligned}
        \det(\Omega_k) &\le \Big(1/N\cdot\text{Tr}\big(\Sigma_0^{-1}\big)+1/N\cdot\sum_{t=0}^{k-1} \text{Tr}(X_tX_t^\top)\Big)^N,
      \end{aligned}
    \end{equation}
    where the trace of the matrix $X_t X_t^\top$ further has
    \begin{equation}
      \begin{aligned}
        \text{Tr}(X_tX_t^\top) &= \text{Tr}\Big(\lambda_t^2A[t:1]^\top H_t^\top V^{-2} H_t A[t:1]\Big)\\
        &\hspace{-5pt}\overset{(2.a)}{\le} 1/\ubar{v}^2 \cdot \sum_{n=1}^N \mathbf{e}_n^\top A[t:1]^\top H_t^\top H_t A[t:1]\mathbf{e}_n\\
        &\hspace{-5pt}\overset{(2.b)}{\le} 1/\ubar{v}^2 \cdot \sum_{n=1}^N \mathbf{e}_n^\top A[t:1]^\top A[t:1]\mathbf{e}_n\\
        &\hspace{-5pt}\overset{(2.c)}{\le} N\cdot\bar{\alpha}/\ubar{v}^2.
      \end{aligned}
    \end{equation}
    Note that $(2.a)$ is due to the assumption $\ubar{v}\cdot\mathbf{I}_M\preceq V\preceq\bar{v}\cdot\mathbf{I}_M$ and $\mathbf{e}_n \in \mathbb{R}^N$ denotes the unit vector; $(2.b)$ follows from the special form of the measurement matrix $H_t$, i.e., each row has only one element equal to one and all others equal to zero; and $(2.c)$ is based on Assumption~\ref{assump:dynamicsBound}. In addition, given that the initialization $\Sigma_0$ has $\ubar{\sigma} \cdot \mathbf{I}_N \;{\preceq}\; \Sigma_0 \;{\preceq}\; \bar{\sigma} \cdot \mathbf{I}_N$, it follows that $\text{Tr}(\Sigma_0^{-1}) \le N/\ubar{\sigma}$ and $\det(\Sigma_0) \le \bar{\sigma}^N$. As a result, we can have
    \begin{equation}
    \begin{aligned}
      &\sqrt{\log\Big({\det(\Omega_k)^{1/2}\det(\Sigma_0)^{1/2}}/{\delta}\Big)} \\
      &=\sqrt{ 1/2\cdot\log\big(\det(\Omega_k)\big) + 1/2\cdot \log\big(\det(\Sigma_0)\big) - \log(\delta)}\\
      &\le \sqrt{N/2}\cdot\sqrt{\log\Big(\frac{\bar{\sigma}/\ubar{\sigma} + \bar{\alpha}\bar{\sigma}\cdot k / \ubar{v}^2}{\delta^{2/N}}\Big)}.
    \end{aligned}
    \end{equation}
    Based on the inequality~\eqref{upperBoundI}, the proof is completed.
  \end{IEEEproof}

  Now, combining Lemmas~\ref{lemma:upperBoundI}--\ref{lemma:upperBoundIII} together with~\eqref{keyIneq}, it has been shown that, {with probability $1 - \delta$}
\begin{equation}
  {\begin{aligned}
  &\big\|\widehat{\bm{\phi}}_k - \bm{\phi}_k\big\|_{\mathcal{D}^{-1/2}_{\Sigma_k},\infty} \\
  &\le \sqrt{N} \cdot \Bigg(\bar{\lambda}B_{k}+C_1 + C_2\sqrt{N}\cdot \sqrt{\log\Big(\frac{\bar{\sigma}/\ubar{\sigma} + \bar{\alpha}\bar{\sigma}\cdot k / \ubar{v}^2}{\delta^{2/N}}\Big)}\Bigg).
\end{aligned}}
\end{equation}
Recall the definition of $\beta_k(\delta)$ in~\eqref{betaK_I}, the inequality in~\eqref{IneqUCBinftyNorm} is proved and so is Proposition~\ref{prop:DUCB_I}.

\subsection{Proof of Theorem~\ref{thm:regretAnalysis_I}}

To facilitate the following proof, let us first introduce a new mapping $\mathbf{a}(\cdot): \mathcal{S}^I \to \mathbb{R}^N$ which translates the positional information \mbox{$\mathbf{p}= \big[\mathbf{p}[1],\mathbf{p}[2],\cdots, \mathbf{p}[I]\big] \in \mathcal{S}^I $} into a $N$-dimensional action vector $\mathbf{a}(\mathbf{p}) \in \mathbb{R}^N$, i.e.,
\begin{align}\label{defAction}
  \mathbf{a}(\mathbf{p}) = \sum_{i=1}^I \mathbf{e}_{s_i},
\end{align}
where each $s_i$ corresponds to the index of the position~$\mathbf{p}[i]$ in the environment $\mathcal{S}$ and $\mathbf{e}_{s_i} \in \mathbb{R}^N$ denotes the unit vector. Now, by the definitions of $\mathbf{p}_k$ and $\mathbf{p}^\star_k$, it can be immediately verified that the vectors $\mathbf{a}(\mathbf{p}_k)$ and $\mathbf{a}(\mathbf{p}^\star_k)$ must have $I$ elements equal to one and all others equal to zero. Further, we denote $\mathcal{A}$ the set of all possibilities of these vectors, i.e.,
\begin{align}
  \mathcal{A} := \{\mathbf{a} \, |\, \mathbf{a} \in \{0,1\}^N, \mathbf{1}^\top \mathbf{a} = I\}.
\end{align}
For simplicity, we abbreviate the above $\mathbf{a}(\mathbf{p}_k)$ and $\mathbf{a}(\mathbf{p}^\star_k)$ to $\mathbf{a}_k \in \mathcal{A}$ and $\mathbf{a}^\star_k \in \mathcal{A}$, subsequently. Based on the definition of $F_k(\cdot)$ as well as the introduced notations, the regret $r_k$ can be expressed as,
\begin{align}\label{lossValue}
  {r_k} = F_k\big(\mathbf{p}^\star_k\big) - F_k(\mathbf{p}_k) = \langle \mathbf{a}^\star_k - \mathbf{a}_k, \bm{\phi}_k\rangle.
\end{align}

To proceed, we show the following lemma which provides an upper bound for the regret~$r_k$ at each time-step $k$.
\begin{lemma}\label{lemma:regretUpperBound}
	Under the conditions in Proposition~\ref{prop:DUCB_I} and let the positional information $\mathbf{a}_k$'s be generated by Algorithm~\ref{algo:UCB}, then it holds with probability at least $1-\delta$ that,
	\begin{align}
		r_k \le 2\sqrt{N} \beta_k(\delta) \cdot\|\mathbf{a}_k\|_{\mathcal{D}_{\Sigma_k}}.
	\end{align}
\end{lemma}
\begin{IEEEproof}
	By the expression of $r_k$ in \eqref{lossValue}, it follows that
	\begin{equation}\label{lossIneq}
  \begin{aligned}
    r_k & =  \langle \mathbf{a}^\star_k, \; \bm{\phi}_k\rangle - \langle \mathbf{a}_k, \; \bm{\phi}_k\rangle \\
    &\hspace{-5pt}\overset{(3.a)}{\le}  \langle \mathbf{a}_k, \; \bm{\mu}_k -  \bm{\phi}_k\rangle \\
    &\hspace{-5pt}\overset{(3.b)}{\le} \|\mathbf{a}_k\|_{\mathcal{D}^{1/2}_{\Sigma_{k}},1} \cdot \|\bm{\mu}_k -  \bm{\phi}_k\|_{\mathcal{D}^{-1/2}_{\Sigma_{k}},\infty}\\
    &\hspace{-5pt}\overset{(3.c)}{\le}2\beta_k(\delta)\cdot \|\mathbf{a}_k\|_{\mathcal{D}^{1/2}_{\Sigma_{k}},1}\\
    &\hspace{-5pt}\overset{(3.d)}{\le}2\sqrt{N}\beta_k(\delta)\cdot \|\mathbf{a}_k\|_{\mathcal{D}_{\Sigma_{k}}},
  \end{aligned}
\end{equation}
where $(3.a)$ is due to $\langle \mathbf{a}^\star_k, \; \bm{\phi}_k\rangle \le \langle \mathbf{a}^\star_k, \; \bm{\mu}_k\rangle\le \langle \mathbf{a}_k, \; \bm{\mu}_k\rangle$~which can be verified by Proposition~\ref{prop:DUCB_I} and the definition of $\mathbf{a}_k$; $(3.b)$ follows from the H\"older's inequality and the fact that $\|\cdot\|_{\mathcal{D}_M, 1}$ and $\|\cdot\|_{\mathcal{D}_M^{-1}, \infty}$ are dual norms; $(3.c)$ is based on the inequality \eqref{IneqUCBinftyNorm} which has been proved previously; and $(3.d)$ is due to Lemma~\ref{lemma:diagNorm}-2). 
\end{IEEEproof}

Based on the above Lemma~\ref{lemma:regretUpperBound}, it is shown that the regret can be upper bounded by $\|\mathbf{a}_k\|_{\mathcal{D}_{\Sigma_{k}}}$. In order to investigate the key term $\|\mathbf{a}_k\|_{\mathcal{D}_{\Sigma_{k}}}$, we next show in the following lemma an intermediate result which can be used to bound $\|\mathbf{a}_k\|_{\mathcal{D}_{\Sigma_{k}}}$.

\begin{lemma}\label{lemma:minTrace}
	Under the conditions in Proposition~\ref{prop:DUCB_I}, it holds,
	\begin{equation}
	\begin{aligned}
		\sum_{k=0}^{K-1} \min&\big\{1, \lambda_k \text{Tr}(Y_k \Sigma_k)\big\} \le 2N\cdot \log\Big(\bar\sigma/\ubar{\sigma}+ K \bar{\sigma}\bar{\alpha}I/ \ubar{v}\Big).
	\end{aligned}		
	\end{equation}
\end{lemma}

\begin{IEEEproof}
	Recall the recursion~\eqref{upsilon_change} of $\Upsilon_k$, the matrix can be also generated as follows,
	\begin{align}\label{UpsilonRecursion_I}
		\Upsilon_{k+1} = \Upsilon_k + \lambda_k A[k:1]^\top Y_k A[k:1].
	\end{align}
	For simplicity, let us further denote  $A[k:1]^\top Y_k A[k:1]$ by~a new matrix $\Xi_k \in \mathbb{R}^N$.~Now, considering determinant of $\Upsilon_k$'s, it then holds that
	\begin{equation}\label{deter}
		\begin{aligned}
			\det(\Upsilon_{k+1}) &= \det\Big(\Upsilon_k^{1/2}\big(\mathbf{I} + \lambda_k\Upsilon_k^{-1/2} \Xi_k \Upsilon_k^{-1/2}\big)\Upsilon_k^{1/2}\Big)\\
			&= \det(\Upsilon_k)\cdot \det\big(\mathbf{I} + \lambda_k \Upsilon_k^{-1/2} \Xi_k \Upsilon_k^{-1/2}\big)\\
			&\hspace{-5pt}\overset{(4.a)}{=}  \det(\Upsilon_k)\cdot \prod_{n=1}^N \Big(1 + \lambda_k \bm\lambda_n(\Upsilon_k^{-1/2} \Xi_k \Upsilon_k^{-1/2})\Big)\\
			&\hspace{-5pt}\overset{(4.b)}{\ge}  \det(\Upsilon_k)\cdot \Big(1 + \sum_{n=1}^N\lambda_k \bm\lambda_n(\Upsilon_k^{-1/2} \Xi_k \Upsilon_k^{-1/2})\Big)\\
			&= \det(\Upsilon_k)\cdot \Big(1 + \lambda_k \text{Tr}(\Upsilon_k^{-1/2} \Xi_k \Upsilon_k^{-1/2})\Big),
		\end{aligned}
	\end{equation}
	where $\bm\lambda_n(\cdot)$ denotes the $n$-th eigenvalue of the matrix in~$(4.a)$ and $(4.b)$ is due to the inequality of arithmetic and geometric means. Based on the cyclic property of the matrix trace and the recursion of $\Sigma_k$ in~\eqref{Kalman_recursion_Sigma}, it follows
  \begin{equation}
    \begin{aligned}
    \text{Tr}(\Upsilon_k^{-1/2} \Xi_k \Upsilon_k^{-1/2}) 
    & = \text{Tr}(Y_k \Sigma_k).
  \end{aligned} 
  \end{equation}
	Therefore, \eqref{deter} can be continued as
	\begin{align}\label{deter2}
		\det(\Upsilon_{k+1}) \ge  \det(\Upsilon_k)\cdot \Big(1 + \lambda_k \text{Tr}(Y_k \Sigma_k)\Big).
	\end{align}

	Now, applying the above inequality~\eqref{deter2} recursively yields
	\begin{align}\label{deter3}
		\det(\Upsilon_{k+1}) \ge  \det(\Upsilon_0)\cdot \prod_{t=0}^k\Big(1 + \lambda_t \text{Tr}(Y_t \Sigma_t)\Big).
	\end{align}
Notice that $\min\{1,x\} \le 2\log(1+x)$ is always true for any non-negative scalar $x\ge 0$, thus one can have that
  \begin{equation}\label{minIneq}
    \begin{aligned}
      \sum_{t = 0}^k \min\big\{1, \lambda_t\text{Tr}(Y_t \Sigma_t)\big\}
      &\le \sum_{t=0}^k 2 \log\big(1+\lambda_t\text{Tr}(Y_t \Sigma_t)\big)\\
      &\le 2\log\Big(\det({\Upsilon_{k+1}})/\det(\Upsilon_0)\Big).
    \end{aligned}
  \end{equation}
  Based on the definition~\eqref{upsilon} of $\Upsilon_k$, it follows that
    \begin{equation}\label{deter4}
    \begin{aligned}
      \det(\Upsilon_{k+1})
      &\le \Big(1/N \cdot \text{Tr}(\Sigma_0^{-1}) + 1/N \cdot \sum_{t=0}^{k}\lambda_t\text{Tr} (\Xi_t) \Big)^N\\
      &\le\Big(1/\ubar{\sigma} + (k+1)\cdot \bar{\alpha}I/ \ubar{v} \Big)^N.
    \end{aligned}
  \end{equation}
  Therefore, combining~\eqref{minIneq} and~\eqref{deter4}, we can have that
  \begin{equation}
  \begin{aligned}
  	&\sum_{t = 0}^k \min\big\{1, \lambda_t\text{Tr}(Y_t \Sigma_t)\big\}\le 2N\cdot \log\Big(\bar\sigma/\ubar{\sigma}+ (k+1) \bar{\sigma}\bar{\alpha}I/ \ubar{v}\Big),
  \end{aligned}  	
  \end{equation}
  which completes the proof.
\end{IEEEproof}

Next, in order to bound $\|\mathbf{a}_k\|_{\mathcal{D}_{\Sigma_{k}}}$ by using the above Lemma~\ref{lemma:minTrace}, we build the connection between $\text{Tr}(H_t^\top V^{-1}H_t \Sigma_t)$ and $\|\mathbf{a}_k\|_{\mathcal{D}_{\Sigma_{k}}}$ as follows.
\begin{lemma}\label{lemma:conn}
  Let the matrices $\Sigma_k$'s be generated by~\eqref{Kalman_recursion_Sigma} and the positional information $\mathbf{a}_k$'s be generated by Algorithm~\ref{algo:UCB}, then the following statements hold for $\forall k \ge 0$,
  \begin{enumerate}
    \item $\text{Tr}(Y_k \Sigma_k) \ge 1/\bar{v} \cdot \|\mathbf{a}_k\|^2_{\mathcal{D}_{\Sigma_{k}}}$;
    \item $\ubar{v}\cdot \|Y_k\|^2_{\Sigma_k} \le \text{Tr}(Y_k \Sigma_k) \le \bar{v}N\cdot \|Y_k\|^2_{\Sigma_k}$.
  \end{enumerate}
\end{lemma}
\begin{IEEEproof}
  Statement 1): Due to specific forms of covariance matrix $V$ and the measurement matrix $H_k$, it can be confirmed that $Y_k$ has to be diagonal and can be expressed as 
  \begin{align}
    Y_k = \sum_{i=1}^I  \sum_{l \in \mathcal{C}^i} 1/v^i \cdot\mathbf{e}_l\mathbf{e}_l^\top,
  \end{align}
  where $\mathcal{C}_k^i$ denotes the $i$-th agent's sensing area at the time $k$; see definition in~\eqref{MeasureMatrix}. Let us introduce a binary variable $\delta^i_k(n)$; let $\delta^i_k(n) = 1$ if the position indexed by~$n$ is in the sensing area $\mathcal{C}_k^i$, and $\delta^i_n = 0$ otherwise. As a direct result,  it holds that
  \begin{align}\label{trace}
    \text{Tr}(Y_k \Sigma_k) = \sum_{n=1}^N \Big(\sigma_n^k\cdot\sum_{i=1}^I \delta_k^i(n)/v_i \Big),
  \end{align}
  where $\sigma_n^k$ denotes the $n$-th diagonal entry of the matrix $\Sigma_k$. Now, let $s_k^i$ be the index of the agent $i$'s position, one can have that $\delta_k^i (s_k^i) = 1$ and therefore,
  \begin{equation}
    \begin{aligned}
       \text{Tr}(Y_k \Sigma_k)  &\ge 1/\bar{v} \cdot \sum_{i=1}^I  \mathbf{e}_{s^i_k}^\top\Sigma_{k}\mathbf{e}_{s^i_k}\\
       &= 1/\bar{v}\cdot \mathbf{a}_k^\top \mathcal{D}_{\Sigma_{k}} \mathbf{a}_k\\
      &=1/\bar{v}\cdot \|\mathbf{a}_k\|^2_{\mathcal{D}_{\Sigma_{k}}},
    \end{aligned}
  \end{equation}
  where the first equality is due to the definition of $\mathbf{a}_k$.

  Statement 2): Based on the equality~\eqref{trace} and the fact that $\delta^i_k(n)$ is a binary variable, it follows that
  \begin{equation}
    \begin{aligned}
      \text{Tr}(Y_k \Sigma_k) &= \sum_{n=1}^N \Big(\sigma_n^k\cdot\sum_{i=1}^I v_i\big(\delta_k^i(n)/v_i\big)^2 \Big)\\
      &\ge \ubar{v}\cdot\sum_{n=1}^N \Big(\sigma_n^k\cdot\sum_{i=1}^I \big(\delta_k^i(n)/v_i\big)^2 \Big)\\
      & = \ubar{v}\cdot\text{Tr}(Y_k \Sigma_k Y_k)\\
      & \ge \ubar{v} \cdot\|Y_k\|^2_{\Sigma_k},
    \end{aligned}
  \end{equation}
  where the last inequality is due to the definition of the matrix norm $\|\cdot\|_{\Sigma_k}$, i.e., $\|Y_k\|^2_{\Sigma_k}$ equals the largest eigenvalue of the matrix $Y_k \Sigma_k Y_k$.

  On the other hand, one can also have that
    \begin{equation}
    \begin{aligned}
      \text{Tr}(Y_k \Sigma_k) 
      &\le \bar{v}\cdot\sum_{n=1}^N \Big(\sigma_n^k\cdot\sum_{i=1}^I \big(\delta_k^i(n)/v_i\big)^2 \Big)\\
      & = \bar{v}\cdot\text{Tr}(Y_k \Sigma_k Y_k)\\
      & \le \bar{v}N \cdot\|Y_k\|^2_{\Sigma_k}.
    \end{aligned}
  \end{equation}
  Therefore, the proof is completed.
\end{IEEEproof}

With the help of the above lemmas, we are in the~position to prove the theorem. By the definition of regret in~\eqref{lossValue}, it is easy to see that the regret $r_k$ has an uniform upper bound, i.e., $r_k \le\bar{\gamma}:= 2\sqrt{I\bar{\alpha}}\cdot\|\mathbf{\phi}_0\|^2$. Based on the above Lemma~\ref{lemma:regretUpperBound}, we can have that
\begin{equation}
  \begin{aligned}
        r_k &\le \min\big\{\bar{\gamma}, \;2\sqrt{N}\beta_k(\delta)\cdot \|\mathbf{a}_k\|_{\mathcal{D}_{\Sigma_{k}}}\big\} \\
    & \le \beta'_k(\delta)\sqrt{N}\cdot \min\big\{1, \;1/\sqrt{\bar{v}}\cdot \|\mathbf{a}_k\|_{\mathcal{D}_{\Sigma_{k}}}\big\},
  \end{aligned}
\end{equation}
where we denote $\beta'_k(\delta) = \max\{\bar{\gamma}, 2\sqrt{\bar{v}}\beta_k(\delta)\}$. According to the definition~\eqref{betaK_I} of the sequence $\{\beta_k(\delta)\}_{k \in \mathbb{N}_+}$, it follows that $\beta'_k(\delta) \le \beta'_{k+1}(\delta)$. Therefore, the cumulative regret has
\begin{equation}\label{cumulativeRegret}
  \begin{aligned}
        &\sum_{k=0}^{K-1} r_k \le \beta'_K(\delta)\sqrt{N}\cdot \sum_{k=0}^{K-1} \min\big\{1, \;1/\sqrt{\bar{v}}\cdot \|\mathbf{a}_k\|_{\mathcal{D}_{\Sigma_{k}}}\big\}\\
        & = \beta'_K(\delta)\sqrt{N}\cdot \Bigg(\sum_{k=0}^{K-1} \mathbbm{1}_{\lambda_k = 1}\cdot\min\big\{1, \;\sqrt{\lambda_k/\bar{v}}\cdot \|\mathbf{a}_k\|_{\mathcal{D}_{\Sigma_{k}}}\big\}\\
        &\hspace{45pt}+\sum_{k=0}^{K-1} \mathbbm{1}_{\lambda_k < 1}\cdot\min\big\{1, \;1/\sqrt{\bar{v}}\cdot \|\mathbf{a}_k\|_{\mathcal{D}_{\Sigma_{k}}}\big\}\Bigg).
  \end{aligned}
\end{equation}
Note that $\mathbbm{1}_{\lambda_k = 1}$ and $\mathbbm{1}_{\lambda_k < 1}$ represent the indicator functions and the last equality is due to the fact that $\lambda_k \le 1, \forall k\ge 0$. Now, let us investigate the two terms in~\eqref{cumulativeRegret} separately. For the first term, by Lemmas~\ref{lemma:minTrace} and~\ref{lemma:conn}, it follows that
\begin{equation}\label{CR1}
  \begin{aligned}
    &\sum_{k=0}^{K-1} \mathbbm{1}_{\lambda_k = 1}\cdot\min\big\{1, \;\sqrt{\lambda_k/\bar{v}}\cdot \|\mathbf{a}_k\|_{\mathcal{D}_{\Sigma_{k}}}\big\}\\
    &\hspace{-5pt}\overset{(5.a)}{\le} \sqrt{K}\cdot \sqrt{\sum_{k=0}^{K-1} \min\big\{1, \;{\lambda_k/\bar{v}}\cdot \|\mathbf{a}_k\|^2_{\mathcal{D}_{\Sigma_{k}}}\big\}}\\
    &\hspace{-5pt}\overset{(5.b)}{\le} \sqrt{K}\cdot \sqrt{\sum_{k=0}^{K-1} \min\big\{1, \;{\lambda_k}\cdot \text{Tr}(Y_k \Sigma_k)\big\}}\\
    &\hspace{-5pt}\overset{(5.c)}{\le} \sqrt{2NK\cdot \log\big(\bar\sigma/\ubar{\sigma}+ K \bar{\alpha}I/ \ubar{v}\big)},
  \end{aligned}
\end{equation}
where $(5.a)$ is due to the inequality of arithmetic and geometric means and the fact that $\sum_{k=0}^{K-1} \mathbbm{1}_{\lambda_k=1} \le K$; $(5.b)$ is based on Lemma~\ref{lemma:conn}-1); and $(5.c)$ is according to Lemma~\ref{lemma:minTrace}. For the second term, it holds that
\begin{equation}\label{CR2}
  \begin{aligned}
    &\sum_{k=0}^{K-1} \mathbbm{1}_{\lambda_k < 1}\cdot\min\big\{1, \;1/\sqrt{\bar{v}}\cdot \|\mathbf{a}_k\|_{\mathcal{D}_{\Sigma_{k}}}\big\}\\
    &\hspace{-5pt}\overset{(6.a)}{\le} \sum_{k=0}^{K-1} \mathbbm{1}_{\lambda_k < 1}\cdot\min\big\{1, \;\sqrt{\bar{v}N}\cdot \|Y_k\|_{{\Sigma_{k}}}\big\}\\
    &\hspace{-5pt}\overset{(6.b)}{=} \sum_{k=0}^{K-1} \mathbbm{1}_{\lambda_k < 1}\cdot\min\big\{1, \;\sqrt{\bar{v}N}\lambda_k/\bar{\lambda}\cdot \|Y_k\|^2_{{\Sigma_{k}}}\big\}\\
    &\hspace{-5pt}\overset{(6.c)}{\le} \sum_{k=0}^{K-1} \mathbbm{1}_{\lambda_k < 1}\cdot\min\big\{1, \;\sqrt{\bar{v}N}\lambda_k/(\bar{\lambda}\ubar{v})\cdot \text{Tr}(Y_k\Sigma_{k})\big\}\\
    &\hspace{-5pt}\overset{(6.d)}{\le}  \lambda'\cdot\sum_{k=0}^{K-1}\min\big\{1, \;\lambda_k\cdot\text{Tr}(Y_k\Sigma_{k})\big\}\\
    &\hspace{-5pt}\overset{(6.e)}{\le}  2N\lambda'\cdot\log\big(\bar\sigma/\ubar{\sigma}+ K \bar{\sigma}\bar{\alpha}I/ \ubar{v}\big),
  \end{aligned}
\end{equation}
where $(6.a)$ is due to the fact that $ \|\mathbf{a}_k\|_{\mathcal{D}_{\Sigma_{k}}} \le \bar{v}N\cdot \|Y_k\|_{{\Sigma_{k}}}$ by Lemma~\ref{lemma:conn}; $(6.b)$ is according to the choice of the weights, i.e., $\lambda_k =  \bar{\lambda}/\|Y_k\|_{\Sigma_k}$ given that $\lambda_k <1$; $(6.c)$ is based on Lemma~\ref{lemma:conn}-2); in~$(6.d)$, we let $\lambda' = \max\{1, \sqrt{\bar{v}N}/(\bar{\lambda} \ubar{v})\}$; and $(6.e)$ is according to Lemma~\ref{lemma:minTrace}.

Now, combining the results obtained in~\eqref{cumulativeRegret} -- \eqref{CR2} as well as the defintion of $\beta_k(\delta)$; see~\eqref{betaK_I}, and let $\bar{\lambda} = \sqrt{N}/B_K$, it yields that
\begin{equation}
  \begin{aligned}
    \sum_{k=0}^{K-1} r_k \le \hspace{2pt}& \beta'_K(\delta)\sqrt{N}\cdot \Big(\sqrt{2NK\cdot \log\big(\bar\sigma/\ubar{\sigma}+ K \bar{\sigma}\bar{\alpha}I/ \ubar{v}\big)} \\
    &+ 2N\alpha'\cdot\log\big(\bar\sigma/\ubar{\sigma}+ K \bar{\sigma}\bar{\alpha}I/ \ubar{v}\big)\Big)\\
    \le \hspace{2pt}& \mathcal{O}\Big(N^{3/2}\sqrt{\log{K}} \cdot\big(\sqrt{NK\log{K}}+ {N}B_K\log{K}\big)\Big)\\
    =\hspace{2pt}&\mathcal{O}\Big(N^{2}\sqrt{K}\log{K} + N^{3/2}B_K\log^{5/2}{K}\Big)\\
    =\hspace{2pt}&\widetilde{\mathcal{O}} \Big(N^{2}\sqrt{K} + N^{5/2}B_K\Big).
  \end{aligned}
\end{equation}
Therefore, the proof is completed.

\subsection{Proof of Proposition~\ref{prop:DUCB_II}}

This proof can be completed by following the similar steps as the one for Proposition~\ref{prop:DUCB_I}, except the main differences in the dynamics of state $\widetilde{\bm\phi}_k$ (due to the different type of disturbance) and the specification of weights $\lambda_k$ and $\omega_k$.

 Taking the dynamics~\eqref{typeII_disturb} into account and following the same steps previously, it can be proved without details that
 \begin{equation}\label{diagError-II}
  \begin{aligned}
    &\big\|\widehat{\bm{\phi}}_k - \overbar{\bm{\phi}}_k\big\|_{\mathcal{D}^{-1/2}_{\Sigma_k},\infty} \\
    &\le  \sqrt{N} \cdot \Big(\big\|\Sigma_0^{-1} ( \widehat{\bm{\phi}}_0 - \widetilde{\bm{\phi}}_0)\big\|_{\Upsilon_k^{-1}} + \lambda_{k-1}\big\| A[k:1]^{-1} \widetilde{\bm\phi}^{k}\big\|_{\Upsilon_k^{-1}}\\
      &\hspace{45pt}+\Big\|\sum_{t=0}^{k-1}\lambda_tA[t:1]^\top H_t^\top V^{-1} \mathbf{n}_t\Big\|_{\Upsilon_k^{-1}}\Big),
  \end{aligned}
\end{equation}
where $\overbar{\bm\phi}_k$ is defined as in~\eqref{tildePhi}. While the first term on the right hand side can be exactly bounded by the previous Lemma~\ref{lemma:upperBoundI}, the last two need specific attentions to obtain the upper bounds.

First, we show in the following lemma that the second term can be indeed bounded by the weight $\lambda_k$.

\begin{lemma}\label{lemma:upperBoundIV}
	Under the conditions in Proposition~\ref{prop:DUCB_II}, there exists a constant $C_3 = \bar{\phi}/\sqrt{\ubar{\alpha}}$, such that
	\begin{align}
		    \lambda_{k-1}\big\| A[k:1]^{-1} \widetilde{\bm\phi}^{k}\big\|_{\Upsilon_k^{-1}} \le C_3 \sqrt{\lambda_{k-1}}.
	\end{align}
\end{lemma}

\begin{IEEEproof}
	By the definition~\eqref{upsilon} of $\Upsilon_k$ and the specification of $\lambda_k = \omega_k = (1/\gamma)^k$, it is verified that ${\Upsilon_k^{-1}} \le 1/(\lambda_{k-1})\cdot \mathbf{I}_N$. Therefore, it holds that
  \begin{equation}
      \begin{aligned}
    \big\|A[k:1]^{-1}\widetilde{\bm\phi}^{k}\big\|^2_{\Upsilon_k^{-1}} 
    &\le  1/(\lambda_{k-1})\cdot \big\|A[k:1]^{-1} \widetilde{\bm\phi}^{k}\big\|^2\\
    & = 1/(\lambda_{k-1})\cdot \| \widetilde{\bm\phi}^{k}\|^2_{A[k:1]^{-\top}A[k:1]^{-1}}\\
    & \le 1/(\ubar{\alpha}\lambda_{k-1})\cdot \| \widetilde{\bm\phi}^{k}\|^2\\
     & \le \bar{\phi}^2/(\ubar{\alpha}\lambda_{k-1}),
     \end{aligned}
  \end{equation}
  where the last two inequalities are due to Assumption~\ref{assump:dynamicsBound} and the condition {$\|\widetilde{\bm\phi}_k\| \le \bar{\phi}$} in Assumption~\ref{assump:stateBound}, respectively. Therefore, the proof is completed.
\end{IEEEproof}

Next, the third term can be handled by applying the result of self-normalized Martingale as in Lemma~\ref{lemma:upperBoundIII}. Nevertheless, to adapt the change of the matrix $\Upsilon_k$, we need to modify the definition of $\Omega_k$ accordingly,
\begin{align}
	\Omega_k := \sum_{t=0}^{k-1} X_t X_t^\top + \lambda_{k-1}^2 \cdot \mathbf{I}_N \in\mathbb{R}^{N \times N},
\end{align}
where $X_t$ is defined as same as before; see equation~\eqref{X}. As a result, we can have that, with probability at least $1-\delta$,
    \begin{align}
      \Big\|\sum_{t=0}^{k-1} X_t\mathbf{n}_t\Big\|_{\Omega_k^{-1}} \le 2\bar{v}^2\cdot\sqrt{\log\Big(\frac{\det(\Omega_k)^{1/2}\cdot\lambda_{k-1}^{-N}}{ \delta}\Big)}.
    \end{align}

    Due to the fact that the sequence $\{\lambda_k\}_{k\in\mathbb{N}_+}$ is increasing {with $\lambda_1 >1$}, it can be proved by following the same steps as before that $ \Omega_k \le\max\{1, 1/\ubar{v}\} \cdot  \lambda_{k-1}\Upsilon_k$ and furthermore,
    \begin{equation}
      \begin{aligned}
         &\Big\|\sum_{t=0}^{k-1} X_t\mathbf{n}_t\Big\|_{\Upsilon_k^{-1}} \le \sqrt{\max\{1, 1/\ubar{v}\} \cdot\lambda_{k-1}} \cdot \Big\|\sum_{t=0}^{k-1} X_t\mathbf{n}_t\Big\|_{\Omega_k^{-1}}\\
         &\le C_2\sqrt{N}\sqrt{\lambda_{k-1}} \cdot \sqrt{\log\Big(\frac{1+\bar{\alpha}/\ubar{v}^2 \lambda_{k-1}^{-2}\cdot\sum_{t=0}^{k-1} \lambda_t^2}{\delta^{2/N}}\Big)},\\
         & = C_2\sqrt{N}\sqrt{\lambda_{k-1}} \cdot \sqrt{\log\Big(\frac{1+\bar{\alpha}/\ubar{v}^2\cdot\sum_{t=0}^{k-1} \gamma^{2(k-t-1)}}{\delta^{2/N}}\Big)},
      \end{aligned}
    \end{equation}
    where $C_2 =  \bar{v}^2 \sqrt{\max\{2,2/\ubar{v}\}}$.

    Now, combining the above inequality together with~\eqref{diagError-II} as well as Lemmas~\ref{lemma:upperBoundI} and~\ref{lemma:upperBoundIV}, one can have that,
 \begin{equation}\label{Ineq_DUCB_II}
  \begin{aligned}
  &\big\|\widehat{\bm{\phi}}_k - \bm{\phi}_k\big\|_{\mathcal{D}^{-1/2}_{\Sigma_k},\infty} \le \sqrt{N} \cdot \Bigg(C_1 + C_3 \sqrt{\lambda_{k-1}}\\
  & + C_2\sqrt{N}\sqrt{\lambda_{k-1}} \cdot \sqrt{\log\Big(\frac{1+\bar{\alpha}/\ubar{v}^2\cdot\sum_{t=0}^{k-1} \gamma^{2(k-t-1)}}{\delta^{2/N}}\Big)}\Bigg).
\end{aligned}
\end{equation}
Therefore, recall the definition of $\beta_k(\delta)$ in~\eqref{betaK_II}, the proof of Proposition~\ref{prop:DUCB_II} is completed.

\subsection{Proof of Theorem~\ref{thm:regretAnalysis_II}}

By applying the notions defined in the proof of Theorem~\ref{thm:regretAnalysis_I}, one  can have that,
\begin{equation}\label{regretBound_II}
\begin{aligned}
  \widetilde{r}_k &= \widetilde{F}_k\big(\mathbf{p}^\star_k\big) - \widetilde{F}_k(\mathbf{p}_k) = \langle \mathbf{a}^\star_k - \mathbf{a}_k, \widetilde{\bm\phi}_k\rangle\\
  &=  \langle \mathbf{a}^\star_k, \; \overbar{\bm{\phi}}_k\rangle - \langle \mathbf{a}_k, \; \overbar{\bm{\phi}}_k\rangle + \langle \mathbf{a}^\star_k - \mathbf{a}_k, \; \widetilde{\bm\phi}_k-\overbar{\bm{\phi}}_k\rangle\\ 
  &\hspace{-5pt}\overset{(7.a)}{\le}  \langle \mathbf{a}_k, \; \bm{\mu}_k - \overbar{ \bm{\phi}}_k\rangle + 2\sqrt{I}\cdot\|\widetilde{\bm\phi}_k-\overbar{\bm{\phi}}_k\|\\
  &\hspace{-5pt}\overset{(7.b)}{\le}  \|\mathbf{a}_k\|_{\mathcal{D}^{1/2}_{\Sigma_{k}},1} \cdot \|\bm{\mu}_k -  \overbar{\bm{\phi}}_k\|_{\mathcal{D}^{-1/2}_{\Sigma_{k}},\infty} + 2\sqrt{I}\cdot\|\widetilde{\bm\phi}_k-\overbar{\bm{\phi}}_k\| \\
    &\hspace{-5pt}\overset{(7.c)}{\le} 2\sqrt{N}\beta_k(\delta)\cdot \|\mathbf{a}_k\|_{\mathcal{D}_{\Sigma_{k}}} + 2\sqrt{I}\cdot\|\widetilde{\bm\phi}_k-\overbar{\bm{\phi}}_k\| ,
\end{aligned}	
\end{equation}
where $(7.a)$ is due to the definitions of $\mathbf{a}^\star_k$ and $\mathbf{a}_k$ as well as the fact that $\langle \mathbf{a}^\star_k, \; \overbar{\bm\phi}_k\rangle \le \langle \mathbf{a}^\star_k, \; \bm{\mu}_k\rangle\le \langle \mathbf{a}_k, \; \bm{\mu}_k\rangle$; $(7.b)$ is from the H\"older's inequality; and $(7.c)$ comes from the inequality~\eqref{Ineq_DUCB_II} which has been proved in the proof of Proposition~\ref{prop:DUCB_II}. It can be seen from the above result that, due to the involvement of all prior disturbances $\bm{\delta}_k$'s in the state $\widetilde{\bm\phi}_k$, the regret $\widetilde{r}_k$ can be no longer bounded by the action term $\|\mathbf{a}_k\|_{\mathcal{D}_{\Sigma_{k}}}$ solely, but needs to consider the extra term which is related to the discrepancy between $\widetilde{\bm\phi}_k$ and $\overbar{\bm{\phi}}_k$. Hence, we next provide an upper bound for $\|\widetilde{\bm\phi}_k-\overbar{\bm{\phi}}_k\|$ with respect to the disturbances $\bm\delta_k$'s.

\begin{lemma}\label{lemma:discrepancy}
	Under the conditions in Proposition~\ref{prop:DUCB_II}, it holds that, for $ 0 \le D \le k$, 
  \begin{equation}
      \begin{aligned}
    &\|\widetilde{\bm\phi}_k - \overbar{\bm{\phi}}_k \|  \le C_4/\lambda_{k-1} + C_5/\lambda_{k-1}\hspace{-4pt} \sum_{t=0}^{k-D-1}\lambda^{t}+ C_6\hspace{-4pt} \sum_{t = k-D}^{k-1} \| \bm\delta_t\|,
  \end{aligned}
  \end{equation}
   where $C_4 = \bar{\phi}\sqrt{\bar{\alpha}}\cdot(1+1/\sqrt{\ubar{\alpha}})/\ubar{\sigma}$, $C_5 = {\bar{\alpha}}\bar{\phi}\cdot(1+1/\sqrt{\ubar{\alpha}})/(\ubar{v}I)$, and $C_6 = \sqrt{\bar{\alpha}/\ubar{\alpha}}$.
\end{lemma}
\begin{IEEEproof}
  Recall the definition~\eqref{tildePhi} of $\overbar{\bm{\phi}}_k$, it follows that
  \begin{equation}\label{disturb}
    \begin{aligned}
      &\overbar{\bm{\phi}}_k - \widetilde{\bm\phi}_k  = A[k:1]\Upsilon_k^{-1} \Big( \Sigma_0^{-1} \big(\widetilde{\bm\phi}_0 - A[k:1]^{-1} \widetilde{\bm\phi}_k\big)\\
      &\hspace{10pt}+\hspace{-5pt}\sum_{t=0}^{k-D-1}\lambda_tA[t:1]^\top H_t^\top V^{-1} H_t\big( \widetilde{\bm\phi}_t - A[k:t+1]^{-1} \widetilde{\bm\phi}_k \big)\\
      &\hspace{10pt}+\hspace{-5pt}\sum_{t=k-D}^{k-1}\lambda_tA[t:1]^\top H_t^\top V^{-1} H_t\big( \widetilde{\bm\phi}_t - A[k:t+1]^{-1} \widetilde{\bm\phi}_k \big)\Big).
    \end{aligned}
  \end{equation}
We next upper bound in order the three terms on the right-hand-side of~\eqref{disturb}.

  \textbf{Terms I}: Due to the fact that $\Upsilon_k^{-1} \le 1/\lambda_{k-1} \cdot \mathbf{I}_N$, it holds
  \begin{equation}\label{51}
    \begin{aligned}
      &\Big\|A[k:1]\Upsilon_k^{-1} \Sigma_0^{-1} \big(\widetilde{\bm\phi}_0 - A[k:1]^{-1} \widetilde{\bm\phi}_k\big)\Big\|\\
       &\le \big\|\widetilde{\bm\phi}_0\big\|_{M_1} + \big\| \widetilde{\bm\phi}_k\big\|_{M_2}.
    \end{aligned}
  \end{equation}
  Note that, in~\eqref{51}, the two matrices $M_1$ and $M_2$ have
  \begin{equation}
    \begin{aligned}
      M_1 :&=\Sigma_0^{-1}\Upsilon_k^{-1}A[k:1]^\top A[k:1]\Upsilon_k^{-1} \Sigma_0^{-1} \\
      &\le \bar{\alpha}/ (\lambda_{k-1} \ubar{\sigma})^{2}\cdot \mathbf{I}_N,
    \end{aligned}
  \end{equation}
  and
  \begin{equation}
    \begin{aligned}
      M_2 :&=A[k\hspace{-1pt}:\hspace{-1pt}1]^{-\top}\Sigma_0^{-1}\Upsilon_k^{-1}A[k\hspace{-1pt}:\hspace{-1pt}1]^\top A[k\hspace{-1pt}:\hspace{-1pt}1]\Upsilon_k^{-1} \Sigma_0^{-1}A[k\hspace{-1pt}:\hspace{-1pt}1]^{-1}\\
      & \le \bar{\alpha}/\big(\ubar{\alpha} (\lambda_{k-1} \ubar{\sigma})^{2}\big)\cdot \mathbf{I}_N.
    \end{aligned}
  \end{equation}
  As a result of $\|\widetilde{\bm\phi}_k\| \le \bar{\phi}$ in Assumption~\ref{assump:stateBound}, we can have that
  \begin{align}\label{IneqDisturbI}
    \Big\|A[k:1]\Upsilon_k^{-1} \Sigma_0^{-1} \big(\widetilde{\bm\phi}_0 - A[k:1]^{-1} \widetilde{\bm\phi}_k\big)\Big\| \le C_4 / \lambda_{k-1},
  \end{align}
  where $C_4 = \bar{\phi}\sqrt{\bar{\alpha}}\cdot(1+1/\sqrt{\ubar{\alpha}})/\ubar{\sigma}$.

  \textbf{Term II}: Following the same path for the analysis of the first term, it can be shown that
  \begin{equation}\label{IneqDisturbII}
    \begin{aligned}
      &\Big\|A[k:1]\Upsilon_k^{-1}\\
      &\boldsymbol{\cdot}\sum_{t=0}^{k-D-1}\lambda_tA[t:1]^\top H_t^\top V^{-1} H_t\big( \widetilde{\bm\phi}_t - A[k:t+1]^{-1} \widetilde{\bm\phi}_k \big)\Big\|\\
      &\le \bar{\alpha}\bar{\phi}/( \ubar{v}I \lambda_{k-1}) \cdot\sum_{t=0}^{k-D-1} \lambda_t + \bar{\alpha}\bar{\phi}/( \sqrt{\ubar{\alpha}} \ubar{v}I \lambda_{k-1}) \cdot\sum_{t=0}^{k-D-1} \lambda_t\\
      &\le C_5 /\lambda_{k-1} \sum_{t=0}^{k-D-1} \lambda_t,
    \end{aligned}
  \end{equation}
  where $C_5 = {\bar{\alpha}}\bar{\phi}\cdot(1+1/\sqrt{\ubar{\alpha}})/(\ubar{v}I)$.

  \textbf{Term III}: Note that, for $\forall k \ge t+1$, we can have
  \begin{equation}\label{telescope}
      \begin{aligned}
     &A[k:t+1]^{-1} \widetilde{\bm\phi}_k - \widetilde{\bm\phi}_t \\
     &\hspace{20pt}=\sum_{s=t}^{k-1} A[s+1:t+1]^{-1} (\widetilde{\bm\phi}_{s+1} - A_{s+1} \widetilde{\bm\phi}_{s}).
    \end{aligned} 
  \end{equation}
  Therefore, it holds that
\begin{equation}\label{IneqDisturbIII}
    \begin{aligned}
      &\Big\|A[k:1]\Upsilon_k^{-1}\\
      &\boldsymbol{\cdot}\sum_{t=k-D}^{k-1}\lambda_tA[t:1]^\top H_t^\top V^{-1} H_t\big(  A[k:t+1]^{-1} \widetilde{\bm\phi}_k - \widetilde{\bm\phi}_t \big)\Big\|\\
      &\hspace{-5pt}\overset{(8.a)}{\le} \sqrt{\bar{\alpha}}\cdot\Big\|\Upsilon_k^{-1}\sum_{t=k-D}^{k-1} \sum_{s=t}^{k-1} \lambda_t  A[t:1]^\top H_t^\top V^{-1} H_t\\
      &\hspace{40pt}\boldsymbol{\cdot}A[s+1:t+1]^{-1} (\widetilde{\bm\phi}_{s+1} - A_{s+1} \widetilde{\bm\phi}_{s})\Big\|\\
      &\hspace{-5pt}\overset{(8.b)}{=} \sqrt{\bar{\alpha}}\cdot\Big\|\sum_{s=k-D}^{k-1} \Upsilon_k^{-1}\sum_{t=k-D}^{s} \lambda_t  A[t:1]^\top H_t^\top V^{-1} H_tA[t:1]\\
      &\hspace{40pt}\boldsymbol{\cdot}A[s+1:1]^{-1} (\widetilde{\bm\phi}_{s+1} - A_{s+1} \widetilde{\bm\phi}_{s})\Big\|\\
      &\hspace{-5pt}\overset{(8.c)}{\le} \sqrt{\bar{\alpha}}\cdot\sum_{s=k-D}^{k-1} \Big\|\Upsilon_k^{-1}\sum_{t=k-D}^{s} \lambda_t  A[t:1]^\top H_t^\top V^{-1} H_tA[t:1]\\
      &\hspace{40pt}\boldsymbol{\cdot}A[s+1:1]^{-1} (\widetilde{\bm\phi}_{s+1} - A_{s+1} \widetilde{\bm\phi}_{s})\Big\|\\
      &\hspace{-5pt}\overset{(8.d)}{\le} \sqrt{\bar{\alpha}}\cdot\sum_{s=k-D}^{k-1} \Big\|A[s+1:1]^{-1} (\widetilde{\bm\phi}_{s+1} - A_{s+1} \widetilde{\bm\phi}_{s})\Big\|\\
      &\hspace{-5pt}\overset{(8.e)}{\le}  \sqrt{\bar{\alpha}/\ubar{\alpha}}\cdot\sum_{s=k-D}^{k-1} \big\| \widetilde{\bm\phi}_{s+1} - A_{s+1} \widetilde{\bm\phi}_{s}\big\|.
    \end{aligned}
  \end{equation}
  Note that $(8.a)$ and $(8.e)$ are due to Assumption~\ref{assump:dynamicsBound}; in $(8.b)$, we exchange the summation indices and apply~\eqref{telescope}; $(8.c)$ follows from the triangle inequality; and $(8.d)$ is due to the fact that
  \begin{equation}
    \begin{aligned}
      &\Upsilon_k^{-1}\sum_{t=k-D}^{s} \lambda_t  A[t:1]^\top H_t^\top V^{-1} H_tA[t:1]\\
      &\le \Upsilon_k^{-1}\Big(\Sigma_0^{-1} \hspace{-2pt}+\hspace{-2pt} \sum_{t=0}^{k-1} \lambda_t  A[t\hspace{-2pt}:\hspace{-2pt}1]^\top H_t^\top V^{-1} H_tA[t\hspace{-2pt}:\hspace{-2pt}1] + \lambda_{k-1}\cdot \mathbf{I}_N\Big)\\
      &= \mathbf{I}_N.
    \end{aligned}
  \end{equation}

  Thus, let $C_6 = \sqrt{\bar{\alpha}/\ubar{\alpha}}$ and based on the upper bounds of the three terms; see inequalities~\eqref{IneqDisturbI},~\eqref{IneqDisturbII} and \eqref{IneqDisturbIII}, the proof of Lemma~\ref{lemma:discrepancy} is completed.
\end{IEEEproof}

In terms of $\|\mathbf{a}_k\|_{\mathcal{D}_{\Sigma_{k}}}$ appearing the regret $\widetilde{r}_k$'s bound~\eqref{regretBound_II}, we apply the same analysis as in the proof of Theorem~\ref{thm:regretAnalysis_I}; see Lemma~\ref{lemma:minTrace} and obtain the following result.

\begin{lemma}\label{lemma:minTrace_II}
	Under the conditions in Proposition~\ref{prop:DUCB_II}, it holds,
	\begin{equation}\label{minTrace_bound_II}
	\begin{aligned}
		&\sum_{k=0}^{K} \min\big\{1, \lambda_k \text{Tr}(Y_k \Sigma_k)\big\} \\
		&\le 2N\cdot \log\Big(\bar\sigma\big(\ubar{\sigma}^{-1} +\lambda_{K}+ \bar{\alpha}I/ \ubar{v}\cdot\sum_{k=0}^{K} \lambda_k \big)\Big).
	\end{aligned}		
	\end{equation}
\end{lemma}

\begin{IEEEproof}
	This proof can be finished by following the~same steps as in the one for  Lemma~\ref{lemma:minTrace}. However, two differences should be noted which are resulted from the distinct definition of $\Upsilon_k$.

	First, under the conditions in this lemma, the recursion of $\Upsilon_k$ follows
	\begin{align}
		\Upsilon_{k+1} = \Upsilon_k + \lambda_k A[k:1]^\top Y_k A[k:1] + (\lambda_k - \lambda_{k-1})\cdot \mathbf{I}_N.
	\end{align}
	Despite the difference as compared to~\eqref{UpsilonRecursion_I},  due to the fact that 
	\begin{align}
		\det\big({\Upsilon_{k+1}}\big) \ge \det\Big({\Upsilon_k + \lambda_k A[k:1]^\top Y_k A[k:1]}\Big),
	\end{align}
	the (in)equalities in~\eqref{deter} are still valid, and so is the subsequent deduction. At last, based on the definition of $\Upsilon_k$ in~\eqref{upsilon}, the final bound in~\eqref{minTrace_bound_II} is obtained by
	    \begin{equation}
    \begin{aligned}
      \det(\Upsilon_{k+1})
      &\le \Big(1/N \cdot \text{Tr}(\Sigma_0^{-1}) + 1/N \cdot \sum_{t=0}^{k}\lambda_t\text{Tr} (\Xi_t) + \lambda_k \Big)^N\\
      &\le\Big(\ubar{\sigma}^{-1} +\lambda_k +  \bar{\alpha}I/ \ubar{v} \cdot \sum_{t=0}^k \lambda_t\Big)^N.
    \end{aligned}
  \end{equation}
\end{IEEEproof}

With the help of the above lemmas, we are ready to prove the sub-linear regret as stated in Theorem~\ref{thm:regretAnalysis_II}. Notice that an uniform upper bound $\bar{\gamma} := 2\sqrt{I} \bar{\phi}$ still exists for the regret $\widetilde{r}_k$. Therefore, it follows from \eqref{regretBound_II} that
  \begin{equation}\label{lossBound}
  \begin{aligned}
    &\widetilde{r}_k \le  \min\big\{\bar{\gamma}, \;2\sqrt{N}\beta_k(\delta)\cdot\|\mathbf{a}_k\|_{\mathcal{D}_{\Sigma_{k}}}\big\}+2\sqrt{I}\cdot\|\widetilde{\bm\phi}_k-\overbar{\bm{\phi}}_k\| \\
     &\le \sqrt{N} {\beta}'_k(\delta) \min\big\{1, \sqrt{\lambda_k/\bar{v}}\cdot \|\mathbf{a}_k\|_{\mathcal{D}_{\Sigma_{k}}}\big\}+ 2\sqrt{I}\cdot\|\widetilde{\bm\phi}_k-\overbar{\bm{\phi}}_k\|,
  \end{aligned}
\end{equation}
where we let $\beta'_k(\delta) := \max\{\bar{\gamma},\,2\sqrt{\bar{v}/\lambda_k} \beta_k(\delta)\}$ in the last~inequality. Therefore, it holds that
\begin{equation}\label{finalIneq}
  \begin{aligned}
    &\sum_{k=0}^{K-1}\widetilde{r}_k\le \sqrt{N} {\beta}'_K(\delta)\cdot\sum_{k=0}^{K-1} \min\big\{1, \;\sqrt{\lambda_k/\bar{v}}\cdot \|\mathbf{a}_k\|_{\mathcal{D}_{\Sigma_{k}}}\big\} \\
    &\hspace{10pt}+ 2\sqrt{I}\cdot\sum_{k=0}^{K-1}\|\widetilde{\bm\phi}_k-\overbar{\bm{\phi}}_k\|\\
     &\hspace{-5pt}\overset{(9.a)}{\le} \sqrt{N} {\beta}'_K(\delta)\cdot \sqrt{K\cdot\sum_{k=0}^{K-1} \min\big\{1, \; \lambda_k\text{Tr}(Y_k \Sigma_k)\big\}}\\
     &\hspace{10pt}+ 2\sqrt{I}\cdot\sum_{k=0}^{K-1}\|\widetilde{\bm\phi}_k-\overbar{\bm{\phi}}_k\|\\
     &\hspace{-5pt}\overset{(9.b)}{\le}N {\beta}'_K(\delta) \sqrt{ 2K}\cdot\log\Big(\bar\sigma\big(\ubar{\sigma}^{-1} +\lambda_{K}+ \bar{\alpha}I/ \ubar{v}\cdot\sum_{k=0}^{K-1} \lambda_k \big)\Big) \\
     &\hspace{10pt} + 2\sqrt{I}C_4 \cdot\sum_{k=0}^{K-1} 1/\lambda_k + 2\sqrt{I}C_5 \cdot\sum_{k=0}^{K-1} 1/\lambda_k \cdot\sum_{t=0}^{k-D-1}\lambda_t \\
     &\hspace{10pt}+2\sqrt{I} C_6\cdot\sum_{k=0}^{K-1} \sum_{t = k-D}^{k-1} \| \bm{\phi}_{t+1} - A_{t+1} \bm{\phi}_{t}\|\\
    &\hspace{-5pt}\overset{(9.c)}{\le}N {\beta}'_K(\delta)\sqrt{ 2K} \cdot\log\Big(\bar\sigma\big(\ubar{\sigma}^{-1} \hspace{-2pt}+\hspace{-2pt}\gamma^{-K}\hspace{-2pt}+\hspace{-2pt} \bar{\alpha}I/ \ubar{v}\cdot\gamma^{-K}/(1-\gamma)\big)\Big) \\
     &\hspace{10pt} + 2\sqrt{I}C_4 /(1-\gamma) + 2\sqrt{I}C_5\cdot (K-D)\gamma^{D+1}/ (1-\gamma) \\
     &\hspace{10pt}+2\sqrt{I} C_6\cdot DB_K,
  \end{aligned}
\end{equation}
where $(9.a)$ is due to the Cauchy-Schwartz inequality and Lemma~\ref{lemma:conn}-1); $(9.b)$ is by Lemmas~\ref{lemma:discrepancy} and~\ref{lemma:minTrace_II}; and $(9.c)$ follows from the specification of $\lambda_k = (1/\gamma)^k$ with $0 < \gamma <1$.

Now, provided that $\gamma = 1 \hspace{-1pt}-\hspace{-1pt} (B_K /K)^{2/3}$; see the condition of Theorem~\ref{thm:regretAnalysis_II}, and letting $D = \lfloor \log(K)/(1\hspace{0pt}-\hspace{0pt}\gamma)\rfloor$, it can be confirmed that $D \le B_K^{-2/3} K^{2/3} \log(K)$ and therefore,
\begin{align}\label{in1}
  2\sqrt{I} C_6\cdot DB_K = \widetilde{\mathcal{O}}\Big(B_K^{1/3} K^{2/3}\Big).
\end{align}
Further, considering that $\log(1/\gamma) \sim 1-\gamma = (B_K /K)^{2/3}$, it holds that $\gamma^D = e^{D\log (\gamma)}\le e^{\log(K)\log(\gamma)/{(1 - \gamma)} } =  \widetilde{\mathcal{O}}(1/K)$, and consequently, 
\begin{equation}\label{in2}
  \begin{aligned}
  &2\sqrt{I}C_4 /(1-\gamma) + 2\sqrt{I}C_5\cdot (K-D)\gamma^{D+1}/ (1-\gamma)\\
   &\sim 1/(1-\gamma) = \widetilde{\mathcal{O}}\Big(B_K^{-2/3}K^{2/3}\Big).
\end{aligned}
\end{equation}
According to the definitions of ${\beta}'_k(\delta)$ and ${\beta}_k(\delta)$, it holds that
\begin{equation}
  \begin{aligned}
    {\beta}'_K(\delta) &\sim \beta_K(\delta)/\sqrt{\lambda_K} \sim N\sqrt{\log\Big(\sum_{t=0}^{K-1} \gamma^{2(K-t-1)} \Big)} \\
    & \le N\sqrt{\log\big({1}/(1 - \gamma) \big)}=  {\mathcal{O}}\Big( N\sqrt{\log(K/B_K)}\Big).
  \end{aligned}
\end{equation}
As a result, one can have that
\begin{equation}\label{in3}
  \begin{aligned}
    &N {\beta}'_K(\delta)\sqrt{ 2K} \cdot\log\Big(\bar\sigma\big(\ubar{\sigma}^{-1} \hspace{-2pt}+\hspace{-2pt}\gamma^{-K}\hspace{-2pt}+\hspace{-2pt} \bar{\alpha}I/ \ubar{v}\cdot\gamma^{-K}/(1-\gamma)\big)\Big)\\
     &\sim  N^2\sqrt{\log(K/B_K)}\cdot \sqrt{K}\cdot \sqrt{K\log( 1/\gamma) + \log\big( 1/(1-\gamma)\big)}\\
     & = \widetilde{\mathcal{O}}\Big( N^2 B_K^{1/3} K^{2/3}\Big).
  \end{aligned}
\end{equation}

At last, combining~\eqref{finalIneq}--\eqref{in2} and~\eqref{in3} arrives at the conclusion in Theorem~\ref{thm:regretAnalysis_II}, i.e., the cumulative regret generated by our algorithm is upper bounded by $\widetilde{R}_K \le \widetilde{\mathcal{O}} \big(N^2B_K^{1/3} K^{2/3}\big)$.

\bibliographystyle{unsrt}
\bibliography{ref}

\begin{thebibliography}{10}

\bibitem{poveda2021robust}
J.~Poveda, M.~Benosman, and R.~Teel, A.and~Sanfelice.
\newblock Robust coordinated hybrid source seeking with obstacle avoidance in
  multi-vehicle autonomous systems.
\newblock {\em IEEE Transactions on Automatic Control}, 2021.

\bibitem{angelico2021source}
B.~Ang{\'e}lico, L.~Chamon, S.~Paternain, A.~Ribeiro, and G.~Pappas.
\newblock Source seeking in unknown environments with convex obstacles.
\newblock In {\em Proceedings of 2021 American Control Conference}, pages
  5055--5061. IEEE, 2021.

\bibitem{li2021source}
T.~Li, B.~Jayawardhana, A.~Kamat, and A.~Kottapalli.
\newblock Source-seeking control of unicycle robots with {3-D} printed flexible
  piezoresistive sensors.
\newblock {\em IEEE Transactions on Robotics}, 2021.

\bibitem{liu2020semi}
W.~Liu, X.~Huo, G.~Duan, and K.~Ma.
\newblock Semi-global stability analysis of source seeking with dynamic sensor
  reading and a class of nonlinear maps.
\newblock {\em International Journal of Control}, pages 1--10, 2020.

\bibitem{ramirez2018stochastic}
E.~Ramirez-Llanos and S.~Martinez.
\newblock Stochastic source seeking for mobile robots in obstacle environments
  via the {SPSA} method.
\newblock {\em IEEE Transactions on Automatic Control}, 64(4):1732--1739, 2018.

\bibitem{azuma2012stochastic}
S.~Azuma, M.~Sakar, and G.~Pappas.
\newblock Stochastic source seeking by mobile robots.
\newblock {\em IEEE Transactions on Automatic Control}, 57(9):2308--2321, 2012.

\bibitem{habibi2016gradient}
J.~Habibi, H.~Mahboubi, and A.~Aghdam.
\newblock A gradient-based coverage optimization strategy for mobile sensor
  networks.
\newblock {\em IEEE Transactions on Control of Network Systems}, 4(3):477--488,
  2016.

\bibitem{rolf2020successive}
E.~Rolf, D.~Fridovich-Keil, M.~Simchowitz, B.~Recht, and C.~Tomlin.
\newblock A successive-elimination approach to adaptive robotic source seeking.
\newblock {\em IEEE Transactions on Robotics}, 37(1):34--47, 2020.

\bibitem{du2021multi}
B.~Du, K.~Qian, H.~Iqbal, C.~Claudel, and D.~Sun.
\newblock Multi-robot dynamical source seeking in unknown environments.
\newblock In {\em Proceedings of 2021 IEEE International Conference on Robotics
  and Automation}, pages 9036--9042. IEEE, 2021.

\bibitem{feiling2018gradient}
J.~Feiling, S.~Koga, M.~Krsti{\'c}, and T.~Oliveira.
\newblock Gradient extremum seeking for static maps with actuation dynamics
  governed by diffusion {PDEs}.
\newblock {\em Automatica}, 95:197--206, 2018.

\bibitem{dougherty2016extremum}
S.~Dougherty and M.~Guay.
\newblock An extremum-seeking controller for distributed optimization over
  sensor networks.
\newblock {\em IEEE Transactions on Automatic Control}, 62(2):928--933, 2016.

\bibitem{li2014cooperative}
Shuai Li, Ruofan Kong, and Yi~Guo.
\newblock Cooperative distributed source seeking by multiple robots: Algorithms
  and experiments.
\newblock {\em IEEE/ASME Transactions on Mechatronics}, 19(6):1810--1820, 2014.

\bibitem{fabbiano2014source}
Ruggero Fabbiano, Carlos~Canudas De~Wit, and Federica Garin.
\newblock Source localization by gradient estimation based on {Poisson}
  integral.
\newblock {\em Automatica}, 50(6):1715--1724, 2014.

\bibitem{brinon2015distributed}
Lara Bri{\~n}{\'o}n-Arranz, Luca Schenato, and Alexandre Seuret.
\newblock Distributed source seeking via a circular formation of agents under
  communication constraints.
\newblock {\em IEEE Transactions on Control of Network Systems}, 3(2):104--115,
  2015.

\bibitem{fabbiano2016distributed}
Ruggero Fabbiano, Federica Garin, and Carlos Canudas-de Wit.
\newblock Distributed source seeking without global position information.
\newblock {\em IEEE Transactions on Control of Network Systems}, 5(1):228--238,
  2016.

\bibitem{atanasov2012stochastic}
Nikolay Atanasov, Jerome Le~Ny, Nathan Michael, and George~J Pappas.
\newblock Stochastic source seeking in complex environments.
\newblock In {\em 2012 IEEE International Conference on Robotics and
  Automation}, pages 3013--3018. IEEE, 2012.

\bibitem{atanasov2015distributed}
Nikolay~A Atanasov, Jerome Le~Ny, and George~J Pappas.
\newblock Distributed algorithms for stochastic source seeking with mobile
  robot networks.
\newblock {\em Journal of Dynamic Systems, Measurement, and Control}, 137(3),
  2015.

\bibitem{zhou1998essentials}
K.~Zhou and J.~Doyle.
\newblock {\em Essentials of Robust Control}, volume 104.
\newblock Prentice hall Upper Saddle River, NJ, 1998.

\bibitem{agarwal2019online}
N.~Agarwal, B.~Bullins, E.~Hazan, S.~Kakade, and K.~Singh.
\newblock Online control with adversarial disturbances.
\newblock In {\em Proceedings of 2019 International Conference on Machine
  Learning}, pages 111--119. PMLR, 2019.

\bibitem{foster2020logarithmic}
D.~Foster and M.~Simchowitz.
\newblock Logarithmic regret for adversarial online control.
\newblock In {\em Proceedings of 2020 International Conference on Machine
  Learning}, pages 3211--3221. PMLR, 2020.

\bibitem{simchowitz2020improper}
M.~Simchowitz, K.~Singh, and E.~Hazan.
\newblock Improper learning for non-stochastic control.
\newblock In {\em Proceedings of 2020 Conference on Learning Theory}, pages
  3320--3436. PMLR, 2020.

\bibitem{pmlr-v117-hazan20a}
E.~Hazan, S.~Kakade, and K.~Singh.
\newblock The nonstochastic control problem.
\newblock In {\em Proceedings of the 31st International Conference on
  Algorithmic Learning Theory}, pages 408--421. PMLR, 2020.

\bibitem{simchowitz2020making}
M.~Simchowitz.
\newblock Making non-stochastic control (almost) as easy as stochastic.
\newblock In {\em Proceedings of the 34th International Conference on Neural
  Information Processing Systems}, pages 18318--18329. PMLR, 2020.

\bibitem{cheung2019learning}
W.~Cheung, D.~Simchi-Levi, and R.~Zhu.
\newblock Learning to optimize under non-stationarity.
\newblock In {\em Proceedings of the 22nd International Conference on
  Artificial Intelligence and Statistics}, pages 1079--1087. PMLR, 2019.

\bibitem{russac2019weighted}
Y.~Russac, C.~Vernade, and O.~Capp{\'e}.
\newblock Weighted linear bandits for non-stationary environments.
\newblock In {\em Proceedings of the 33rd International Conference on Neural
  Information Processing Systems}, pages 12040--12049, 2019.

\bibitem{zhao20a}
Peng Zhao, Lijun Zhang, Yuan Jiang, and Zhi-Hua Zhou.
\newblock A simple approach for non-stationary linear bandits.
\newblock In Silvia Chiappa and Roberto Calandra, editors, {\em Proceedings of
  the Twenty Third International Conference on Artificial Intelligence and
  Statistics}, volume 108 of {\em Proceedings of Machine Learning Research},
  pages 746--755, 2020.

\bibitem{ding2022robust}
Qin Ding, Cho-Jui Hsieh, and James Sharpnack.
\newblock Robust stochastic linear contextual bandits under adversarial
  attacks.
\newblock In {\em International Conference on Artificial Intelligence and
  Statistics}, pages 7111--7123. PMLR, 2022.

\bibitem{bogunovic2021stochastic}
Ilija Bogunovic, Arpan Losalka, Andreas Krause, and Jonathan Scarlett.
\newblock Stochastic linear bandits robust to adversarial attacks.
\newblock In {\em International Conference on Artificial Intelligence and
  Statistics}, pages 991--999. PMLR, 2021.

\bibitem{he2022nearly}
Jiafan He, Dongruo Zhou, Tong Zhang, and Quanquan Gu.
\newblock Nearly optimal algorithms for linear contextual bandits with
  adversarial corruptions.
\newblock In Alice~H. Oh, Alekh Agarwal, Danielle Belgrave, and Kyunghyun Cho,
  editors, {\em Advances in Neural Information Processing Systems}, 2022.

\bibitem{li2019boundedness}
W.~Li, Z.~Wang, D.~Ho, and G.~Wei.
\newblock On boundedness of error covariances for {Kalman} consensus filtering
  problems.
\newblock {\em IEEE Transactions on Automatic Control}, 65(6):2654--2661, 2019.

\bibitem{battistelli2014kullback}
G.~Battistelli and L.~Chisci.
\newblock Kullback--{Leibler} average, consensus on probability densities, and
  distributed state estimation with guaranteed stability.
\newblock {\em Automatica}, 50(3):707--718, 2014.

\bibitem{battistelli2014consensus}
G.~Battistelli, L.~Chisci, G.~Mugnai, A.~Farina, and A.~Graziano.
\newblock Consensus-based linear and nonlinear filtering.
\newblock {\em IEEE Transactions on Automatic Control}, 60(5):1410--1415, 2014.

\bibitem{cattivelli2010diffusion}
F.~Cattivelli and A.~Sayed.
\newblock Diffusion strategies for distributed {Kalman} filtering and
  smoothing.
\newblock {\em IEEE Transactions on automatic control}, 55(9):2069--2084, 2010.

\bibitem{yang2020adversarial}
Lin Yang, Mohammad~Hassan Hajiesmaili, Mohammad~Sadegh Talebi, John~CS Lui,
  Wing~Shing Wong, et~al.
\newblock Adversarial bandits with corruptions: Regret lower bound and
  no-regret algorithm.
\newblock In {\em NeurIPS}, 2020.

\bibitem{olfati2005consensus}
R.~Olfati-Saber and J.~Shamma.
\newblock Consensus filters for sensor networks and distributed sensor fusion.
\newblock In {\em Proceedings of the 44th IEEE Conference on Decision and
  Control}, pages 6698--6703. IEEE, 2005.

\end{thebibliography}
\end{document}